\documentclass[%
 preprint,
 amssymb,
 aps,
]{revtex4-2}

\usepackage{graphicx}
\usepackage{dcolumn}
\usepackage{bm}
\usepackage{xcolor}
\colorlet{RED}{red}
\usepackage{graphicx} 
\usepackage{amsmath}
\allowdisplaybreaks

\DeclareMathOperator{\sech}{sech}

\begin{document}


\title{{Superregular Breathers in the Sasa--Satsuma Equation: Spectral Conditions and Modulation Instability}}
\author{Tongtong Lin}
\author{Huanhe Dong}
\author{Yong Zhang}%
 \email{zhangyong039@sdust.edu.cn}
\affiliation{%
 College of Mathematics and Systems Science,\\ Shandong University of Science and Technology, Qingdao, 266590, China.
}%


%

\date{\today}

\begin{abstract}

{In this work, we study the superregular breathers (SRBs) of the Sasa--Satsuma (SS) equation by combining a Darboux construction with a linear stability calculation. The reduction symmetry of the scalar SS Lax pair makes the fundamental SRB construction degenerate. The paired quasi-Akhmediev branches nevertheless yield second-order SRBs in the focusing sector $n|_{\gamma=0}<0$; neither the defocusing sector nor the focusing sector $n|_{\gamma=0}\geq0$ supports this construction. Linearization about the same plane-wave background gives the modulation-instability gain and, on the admissible branches, reproduces its relation to the group-velocity difference. The carrier wavenumber $b$ enters through the background rather than as a coefficient of the SS equation, and changes the spectral roots and instability interval. In the formal NLS limit, the result reduces to the familiar NLS gain. Numerical evolution of finite-width periodic perturbations produces the predicted second-order profile over the displayed propagation interval.}

\end{abstract}

\maketitle

\section{Introduction}
Breathers are a class of nonlinear excitations that are spatially localized while exhibiting periodic temporal oscillations. Such wave structures have been observed in a wide range of physical systems, including optical waveguides and photonic crystal fibers \cite{8PhysRevLett,9eisenberg2003,33Dudley}, water wave tanks \cite{34PhysRevLett}, Josephson junctions \cite{10APS,11PhysRevLett}, antiferromagnets \cite{12PhysRevLett}, and micromechanical systems \cite{13PhysRevLett}. 
 These excitations serve as fundamental building blocks for more complex nonlinear structures, namely, nonlinear superpositions of breathers, which are closely associated with a variety of important nonlinear phenomena, including rogue waves \cite{14AKHMEDIEV,15AKHMEDIEV,16PhysRevA} and higher-order modulation instability (MI) \cite{17PhysRevLett}. The MI, describing the exponential amplification of long-wavelength perturbations imposed on a continuous-wave background, provides a unified framework for understanding the emergence of various breather solutions \cite{Benjamin-Feir-1967,38ZAKHAROV}. However, the conventional higher-order MI mechanism has an intrinsic limitation: it requires an initial perturbation that is strictly periodic over the entire infinite spatial domain. This limitation can be overcome by superregular breather (SRB) solutions \cite{21Liu,22PhysRevLett}, which arise from a fundamentally different MI scenario.

Unlike the conventional Akhmediev breather (AB), which is localized in time but periodic in space, SRBs are localized in both space and time. This spatial localization originates from their distinct excitation mechanism: an SRB is generated by a small localized perturbation imposed on the continuous-wave background, which excites a pair of quasi-AB solutions propagating in opposite directions with different group velocities \cite{37PhysRevLett}. Unlike the strictly spatially periodic AB solution, each quasi-AB solution possesses a finite-width envelope perturbation superimposed on the AB background and is therefore spatially localized rather than infinitely extended. When two such quasi-AB solutions have the same period and envelope width, their nonlinear superposition gives rise to a higher-order localized structure, namely, the SRB solution. This localized excitation mechanism removes the requirement of an infinitely extended periodic perturbation and provides a more realistic framework for understanding the nonlinear development of MI.

In recent years, SRB solutions have been extensively investigated in the scalar nonlinear Schrödinger (NLS) equation \cite{20PhysRevX}, which provides a fundamental model for describing the propagation of ultrashort pulses in optical fibers, Bose–Einstein condensates, water waves, and plasma physics \cite{1kivshar2003,2PhysRevE,3dalfovo1999,4Benney1967,5zakharov1972}. Beyond the scalar NLS equation, investigations of SRB solutions have been extended to other integrable systems, including the self-induced transparency system \cite{41REN}, the derivative NLS equation \cite{39PhysRevE, XuHeWang2011} and the nonautonomous higher-order NLSE \cite{42ren}, where the relationship between MI and SRB dynamics has been established.
More recently, this research has been generalized to coupled wave systems, such as the Manakov system \cite{22PhysRevLett}. In all these systems, the discrete spectrum of the associated scattering problem plays a crucial role in determining the structure and dynamics of SRB solutions.
{For the SS equation, binary Darboux transformations can be formulated directly from its reduction structure \cite{NimmoYilmaz2015}, and high-order rational localized waves have been constructed by bilinear and hierarchy-reduction methods \cite{FengEtAl2022}. These results provide an appropriate basis for examining the spectral conditions of SRB branches in the SS setting.}
Nevertheless, most existing investigations have focused on the standard NLS framework or its lower-order integrable extensions, whereas the influence of higher-order dispersive and nonlinear effects on SRB excitation remains largely unexplored.

{The SS equation is a higher-order integrable extension of the NLS equation \cite{30Sasa}. Its third-order and derivative nonlinear terms change both the background spectral problem and the linearized modulation spectrum. We use this equation to isolate how these terms, together with the carrier wavenumber of the selected background, alter the SRB conditions.}

The SS equation is given by
\begin{equation}
	\label{eq:eq0}
	\mathrm{i} q_t+\frac{1}{2}q_{xx}+\sigma\left\lvert q \right\rvert^2q+\frac{\mathrm{i}}{6\epsilon}(q_{xxx}+6\sigma (\left\lvert q \right\rvert^2q)_x-3\sigma (\left\lvert q \right\rvert^2)_xq)=0,
\end{equation}
where $\sigma=\pm1$ represents the focusing and defocusing cases, respectively. To facilitate the subsequent analysis, we perform a variable transformation \cite{30Sasa} and obtain
\begin{equation}
	\label{eq:eq}
	q_t+q_{xxx}+6\sigma (\left\lvert q \right\rvert^2q)_x-3\sigma (\left\lvert q \right\rvert^2)_xq=0,
\end{equation}
{Here $t$ is the retarded time, $x$ is the propagation coordinate, and $q$ is the complex envelope. Equation~\eqref{eq:eq} follows from Eq.~\eqref{eq:eq0} by the stated variable transformation. In the normalized equation, the higher-order derivative terms modify the NLS balance; the question considered below is how they affect the spectral and instability conditions for SRB branches.}

{The parameter $\epsilon$ in Eq.~\eqref{eq:eq0} sets the relative scale of the higher-order SS contribution; the NLS model is recovered formally when $\epsilon\to\infty$. Equation~\eqref{eq:eq} itself contains no $b$. In the subsequent construction, $b$ is introduced through the choice of the plane-wave seed $q_0$ and the associated gauge of the Lax pair; in $q_0$, it is the spatial phase gradient (carrier wavenumber). It is therefore a parameter of the selected background and gauge choice, not a coefficient of Eq.~\eqref{eq:eq} or an additional material coefficient. Through this background, it changes the spectral roots, drift velocities, and the MI condition, so the SS instability condition differs from the zero-background-wavenumber NLS benchmark.}

{In this paper, we construct exact SS solutions using a Darboux transformation \cite{Matveev1992DarbouxTA,NimmoYilmaz2015}, derive their group and phase velocities, and identify the corresponding spectral branches. We then linearize about the same background to test the MI criterion independently. Finally, we compare the exact second-order profile with numerical evolution from its localized initial perturbation.}

\section{Spectral problem and Darboux transformation}

The SS equation is an integrable higher-order extension of the NLS equation and admits a Lax pair representation. This spectral structure provides the foundation for constructing exact nonlinear excitations and analyzing their eigenvalue characteristics. In this section, we introduce the Lax pair associated with the SS equation and employ the DT to derive breather solutions on a continuous-wave background.
This formulation enables us to characterize the spectral properties of SRB solutions and analyze their excitation mechanism in the following sections.

The SS equation considered in this work is written as Eq. \eqref{eq:eq}.
Here, $q(x,t)$ denotes the complex-valued envelope field and $\sigma=\pm1$ corresponds to the focusing and defocusing cases, respectively.

The integrability of Eq. \eqref{eq:eq} is characterized by the compatibility condition of the associated Lax pair
\begin{align}
	\label{eq:lax}
     \mathbf{\Phi}_x&=\mathbf{U}\mathbf{\Phi},\quad  \mathbf{\Phi}_t=\mathbf{V}\mathbf{\Phi},\\\nonumber
     \mathbf{U}&=\mathrm{i}\lambda(\sigma_3+\mathbf{I}_3)+\mathbf{Q},\\\nonumber
     \mathbf{V}&=4 \mathrm{i} \lambda^3(\sigma_3+\mathbf{I}_3)+4 \lambda^2\mathbf{Q}+\lambda\mathbf{V}_1 +\mathbf{V}_0,
\end{align}
where $\mathbf{\Phi}$ is the associated eigenfunction and the compatibility condition $\mathbf{\Phi}_{xt}=\mathbf{\Phi}_{tx}$ recovers Eq. \eqref{eq:eq}
with
\begin{align*}
	\label{eq:lax1}
	\mathbf{Q}&=\begin{bmatrix} 0 &\mathrm{i} \sigma q^* & \mathrm{i} \sigma q \\ \mathrm{i} \sigma q & 0&0\\\mathrm{i} \sigma q^*&0&0 \end{bmatrix},\quad\sigma_3=\begin{bmatrix} 1 &0 & 0 \\ 0 & -1&0\\0&0&-1 \end{bmatrix},\\\nonumber
	\mathbf{V}_1&=2 \mathrm{i} \left(\sigma_3\left(\mathbf{Q}^2-\mathbf{Q}_x\right)+b^2 \mathbf{I}_3\right),\\\nonumber
	\mathbf{V}_0&=\left(\mathbf{Q}_x\mathbf{Q}-\mathbf{Q}\mathbf{Q}_x\right)+2\left(\mathbf{Q}^3-\mathbf{Q}_{xx}\right).
\end{align*}
{Here $q^*$ is the complex conjugate of $q$, $\mathbf{I}_3$ is the $3\times3$ identity matrix, and $b\in\mathbb{R}$ is the carrier wavenumber of the selected plane-wave background (after the corresponding gauge choice). To keep the two background branches explicit, we use $b_1=b$ and $b_2=-b$ below.}

To construct exact solutions, we employ the DT associated with the above Lax pair. The DT generates new solutions from a given seed solution while preserving the integrable structure. Since breather solutions correspond to pairs of complex-conjugate spectral parameters, a single DT transformation does not preserve the required reduction symmetry. Therefore, the lowest-order breather solution is obtained through a second-order DT \cite{19LING,NimmoYilmaz2015} (in Appendix)
\begin{equation}
	\label{eq:solution}
	q_1=q_0 \left( 1 - 2\mathrm{i}\chi_i (1 + E(\chi)) (q_a + q_b) \right),
\end{equation}
where the plane wave solution 
\begin{equation*}
	q_0 = a\, \text{exp}({\mathrm{i}\theta_1}),\quad \theta_1 = b \left(x + (b^2 - 6\sigma a^2) t \right),
\end{equation*}
and {$E(\chi) =\sum_{i=1}^{2}\frac{a^2}{(\chi+b_i)(\chi^*+b_i)}$, with $b_1=b$ and $b_2=-b$.}  
The characteristic equation \eqref{eq:charactereq}, we obtain the spectral parameters
\begin{equation*}
	\lambda = \frac{1}{2} \left( \chi - \frac{2\sigma a^2\chi}{\chi^2 - b^2} \right).
\end{equation*}
 Moreover, the $\chi_i$ is the imaginary part of $\chi$, where $\chi$ is the root of the equation
\begin{equation}
	\label{eq:eigenfunction}
	1+\sigma\sum_{i=1}^{2}\frac{a^2}{(\sigma\chi+b_i)(\sigma(\chi+\alpha)+b_i)}=0,
\end{equation}
where $\alpha = \omega + 2\mathrm{i}\gamma$. The parameters $\omega$ and $\gamma$ are the period and the width of the breather respectively.


The solutions of Eq.~\eqref{eq:solution} can evolve into three types of localized structures, the ABs \cite{23akhmediev}, which are localized in $t$ but periodic in $x$; the Kuznetsov-Ma breather \cite{24Kuznetsov, 25ma1979}, which is localized in $x$ but periodic in $t$; and the Peregrine soliton \cite{26Peregrine}, which is localized in both $x$ and $t$. 

The $q_a$ in Eq.\eqref{eq:solution} is defined as follows
\begin{equation*}
	q_a=\frac{F({\chi}) \left( e^{\tau} + e^{-\mathrm{i}\Lambda} \right) 
		+ F({\hat{\chi}}) \left( e^{-\tau} + e^{\mathrm{i}\Lambda} \right)}
	{(1+\mu(\chi)) e^{\tau} + \frac{1+\mu(\hat{\chi})}{ e^{\tau} }
		+ (1+\nu(\chi^*)) e^{\mathrm{i}\Lambda} + \frac{1+\nu(\chi)}{ e^{\mathrm{i}\Lambda}}} ,
	\end{equation*}
	where
	\begin{align*}
		\tau &= 2\gamma (X - V_g T),\quad\Lambda = \omega (X - V_p T),\\\nonumber
	\mu(\chi)&=\sum_{i=1}^{2}\frac{a^2}{(\chi+b_i)(\chi^*+b_i)},\quad F(\chi)=\frac{1}{\chi+b},\\\nonumber \nu(\chi)&=\sum_{i=1}^{2}\frac{a^2}{(\chi+b_i)(\hat{\chi}^*+b_i)},\quad\tilde{F}(\chi)=\frac{1}{\chi-b},
\end{align*}
and the $\hat{\chi}=\chi+\alpha$. For the specific form of $q_b$, see the Appendix \eqref{eq:qb}.

Here, the shifts $X = x - x_{0}$ and $T = t - t_{0}$ involve the offsets $x_{0}$ and $t_{0}$, which fix the breather's position in space and time, respectively. 
While $V_g$ and $V_p$ denote the group and phase velocities, respectively:
\begin{align}
	\label{eq:VgVp}
	V_g =& (4 a^2 - b^2) -\frac{\omega}{\gamma}( 2\lambda_r\chi_i + 2\lambda_i \chi_r + \lambda_i\omega)\\\nonumber&- 4\lambda_r(\chi_r + \omega)+ 4\lambda_i(\chi_i + \gamma),\\\nonumber
	V_p =& (4 a^2- b^2) + \frac{8\gamma}{\omega}(\lambda_r\chi_i + \lambda_r\gamma + \lambda_i\chi_r)\\\nonumber& - 2\lambda_r(2\chi_r+\omega) + 4\lambda_i(\chi_i+2\gamma).
\end{align}
These analytical expressions for the group and phase velocities will be used below to characterize the propagation properties of SRB solutions and establish their connection with the MI dynamics.

\section{Spectral Analysis}

The spectral parameters obtained from the DT determine the structure and propagation characteristics of breather solutions. Therefore, the existence and classification of AB, quasi-AB, and SRB solutions can be analyzed through the distribution of the corresponding eigenvalues. In this section, we investigate the spectral conditions for these nonlinear excitations and identify the parameter regimes in which they exist.
{The solution family, parameterized by $(a,b_j,\gamma,\omega)$ in Eq.~\eqref{eq:solution}, describes a breather-type wave structure that periodically amplifies and decays on a continuous-wave background. Along the spatial direction, the solution has an oscillation period $2\pi/\omega$ and an envelope width $1/(2\gamma)$. Its propagation is governed by the group velocity $V_g(\chi)$ and phase velocity $V_p(\chi)$. We use $\gamma=0$ only for the strictly periodic AB limit when the relevant root admits it; a small nonzero $\gamma$ gives a finite envelope width and is referred to as a quasi-AB.}

To determine the admissible parameter region for SRB solutions, we must first analyze the roots of Eq.\eqref{eq:eigenfunction}
\begin{equation}
	\label{eq:eigenvalue}
	\chi_{1, 2}=-\frac{\alpha}{2}\pm\sqrt{m+\sqrt{n}},
	\chi_{3, 4}=-\frac{\alpha}{2}\pm\sqrt{m-\sqrt{n}},
\end{equation}
where $m=\frac{1}{4}\alpha^2+(b^2-\sigma a^2)$ and $n=(\alpha^2-4\sigma a^2)b^2+a^4$

\begin{figure}[htbp]   
	\centering         
	\includegraphics[width=1\linewidth]{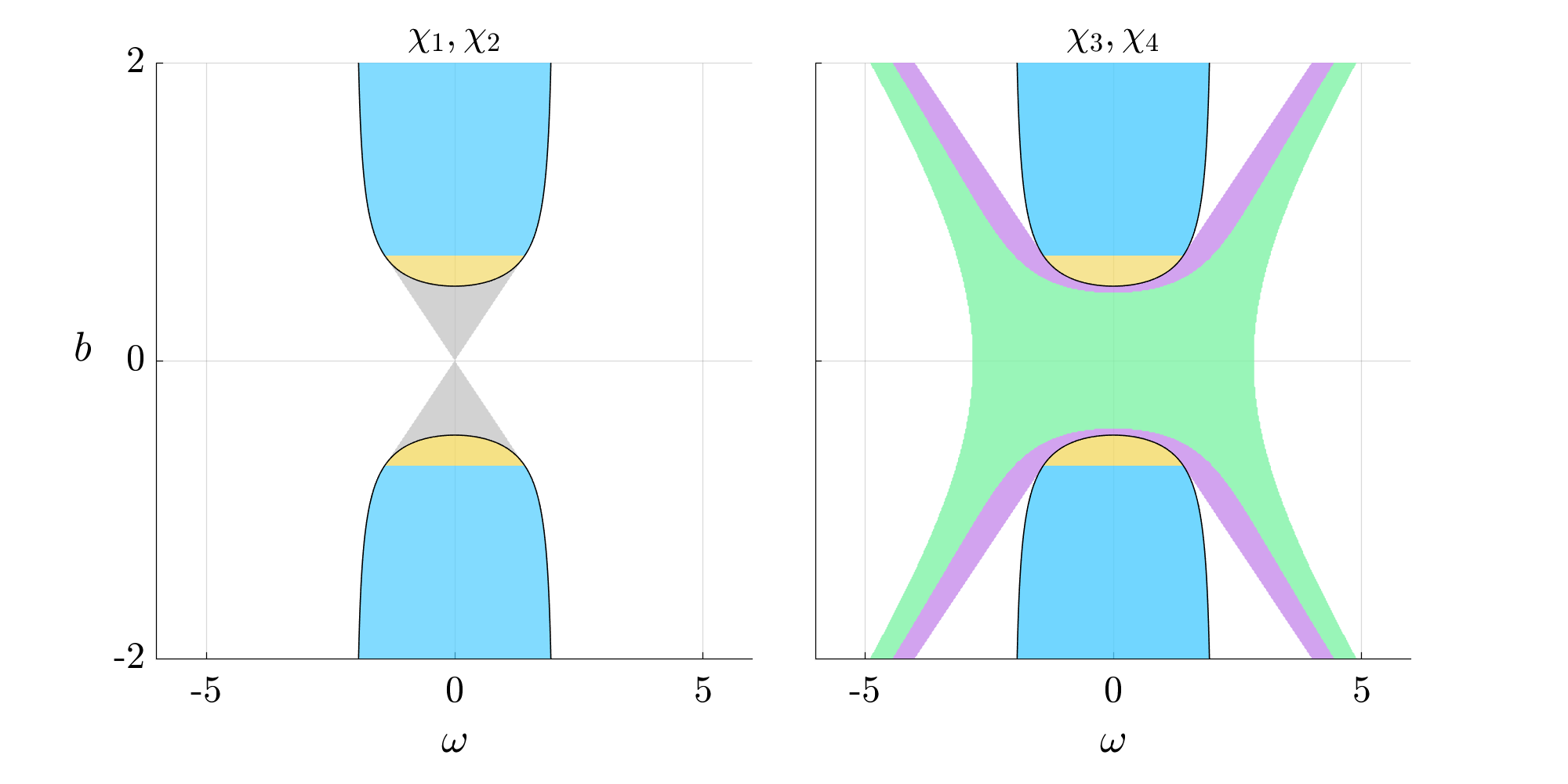}
\caption{{Existence diagram in the focusing regime. Blue: four-petal AB parameters; yellow: twisted-AB parameters; purple and green: the two quasi-AB regimes illustrated in Fig.~\ref{fig:zhunABs}; gray: parameters for which the breather branches considered here are not admissible.}}  
	\label{fig:myLabel}        
\end{figure}

Expanding Eq.\eqref{eq:eigenvalue} to first order with respect to the small parameter $\gamma$ ($\gamma^2 \ll 1$) and separating real and imaginary parts gives
\begin{align}
	\label{eq:chiri}
	\chi_{1, 2i}&=\mathcal{O}(\gamma^2)-\gamma\pm\chi_i^+,\,\,\, \chi_{1, 2r}=\mathcal{O}(\gamma^2)-\frac{\omega}{2}\pm\rho^+,\\\nonumber
	\chi_{3, 4i}&=\mathcal{O}(\gamma^2)-\gamma\pm\chi_i^-,\,\,\, \chi_{3, 4r}=\mathcal{O}(\gamma^2)-\frac{\omega}{2}\pm\rho^-,
\end{align}
where $\chi_i^{\pm}$ and $\rho^{\pm}$ are the imaginary and real parts of the complex parameters when $\gamma = 0$.\\
Namely,
\begin{equation}
	\label{eq:shixu}
	\mathrm{i}\chi_i^{(\mp)}+\rho^{(\mp)}=\left.\sqrt{m\mp\sqrt{n}}\right\lvert_{\gamma=0}.
\end{equation}
Combining Eqs.~\eqref{eq:eigenvalue} and \eqref{eq:shixu}, we obtain the admissible parameter regions for AB, quasi-AB, and SRB solutions, as shown in Fig.~\ref{fig:myLabel}.
{In Fig.~\ref{fig:myLabel}, the blue and yellow regions support four-petal and twisted AB solutions, respectively. The purple and green regions are quasi-AB regimes: in the small-$\gamma$ limit, the relevant $\lambda$ branches become purely imaginary and do not yield the strictly periodic ABs used in the blue and yellow regions.}

{Because the breather spectral parameter $\lambda$ is generally complex, the scalar SS reduction requires a two-fold DT to preserve the required symmetry. The fundamental SRB construction degenerates for the relevant $\chi_3$ and $\chi_4$ branch. Thus, unlike the scalar NLS benchmark, for which a fundamental SRB can occur, the present scalar SS construction yields no fundamental SRB; the nondegenerate SRB states studied below are second order. The defocusing scalar SS case likewise yields no SRB, consistent with Ref.~\cite{22PhysRevLett}.}


In view of the above, we present exact solutions for the purple and green regions, while for the blue and yellow regions we provide both exact solutions and numerical simulations for the ABs and SRBs.
\begin{figure}[htbp]
	\centering
	\noindent\begin{minipage}{\linewidth}
		\begin{minipage}[c]{0.01\linewidth}  
			\centering
			\footnotesize (a)
		\end{minipage}
		\hfill
		\begin{minipage}[c]{0.43\linewidth} 
			\centering
			\includegraphics[width=\linewidth]{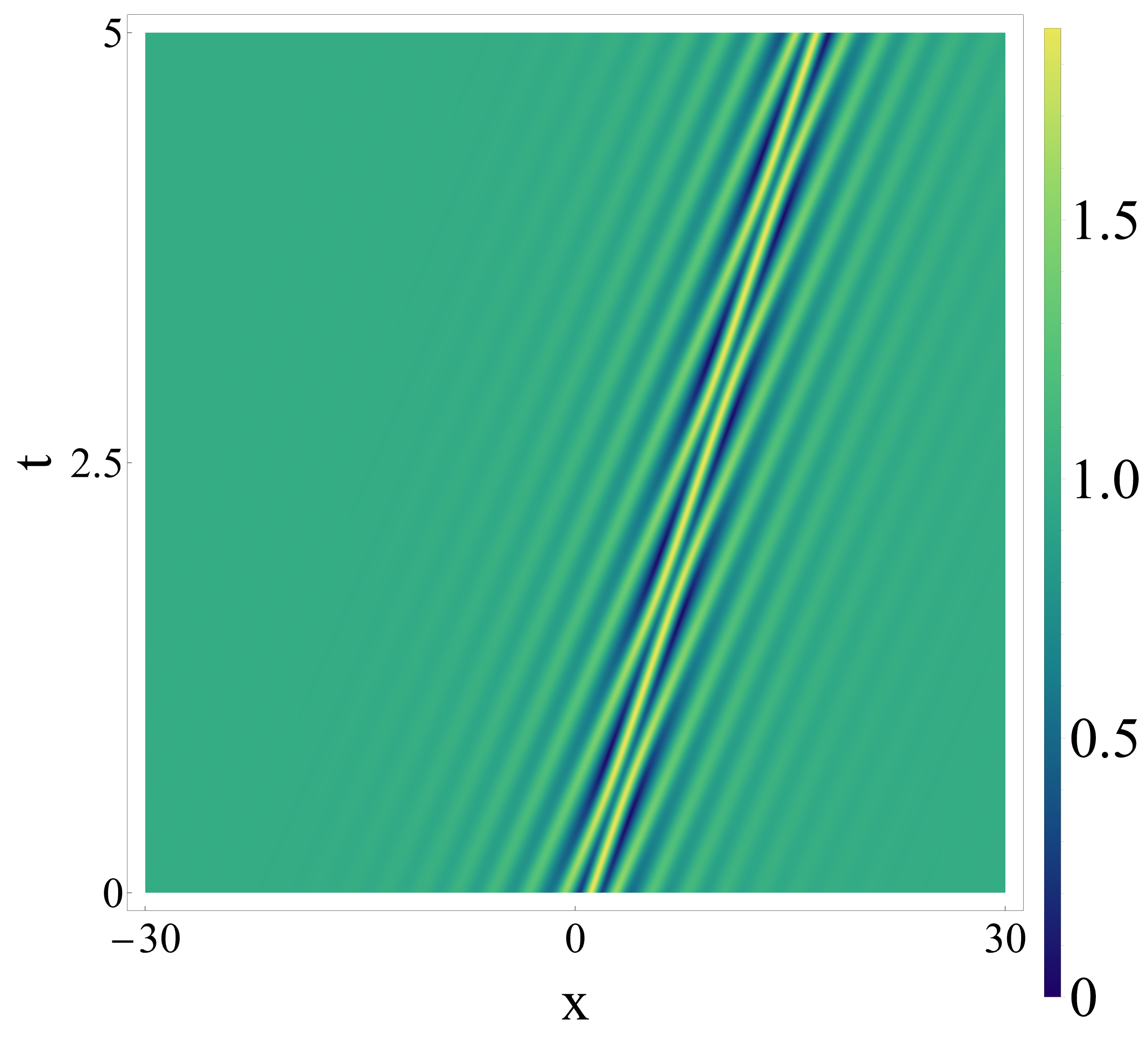}
		\end{minipage}
		\hfill
		\begin{minipage}[c]{0.01\linewidth}  
			\centering
			\footnotesize (c)
		\end{minipage}
		\hfill
		\begin{minipage}[c]{0.43\linewidth}  
			\centering
			\includegraphics[width=\linewidth]{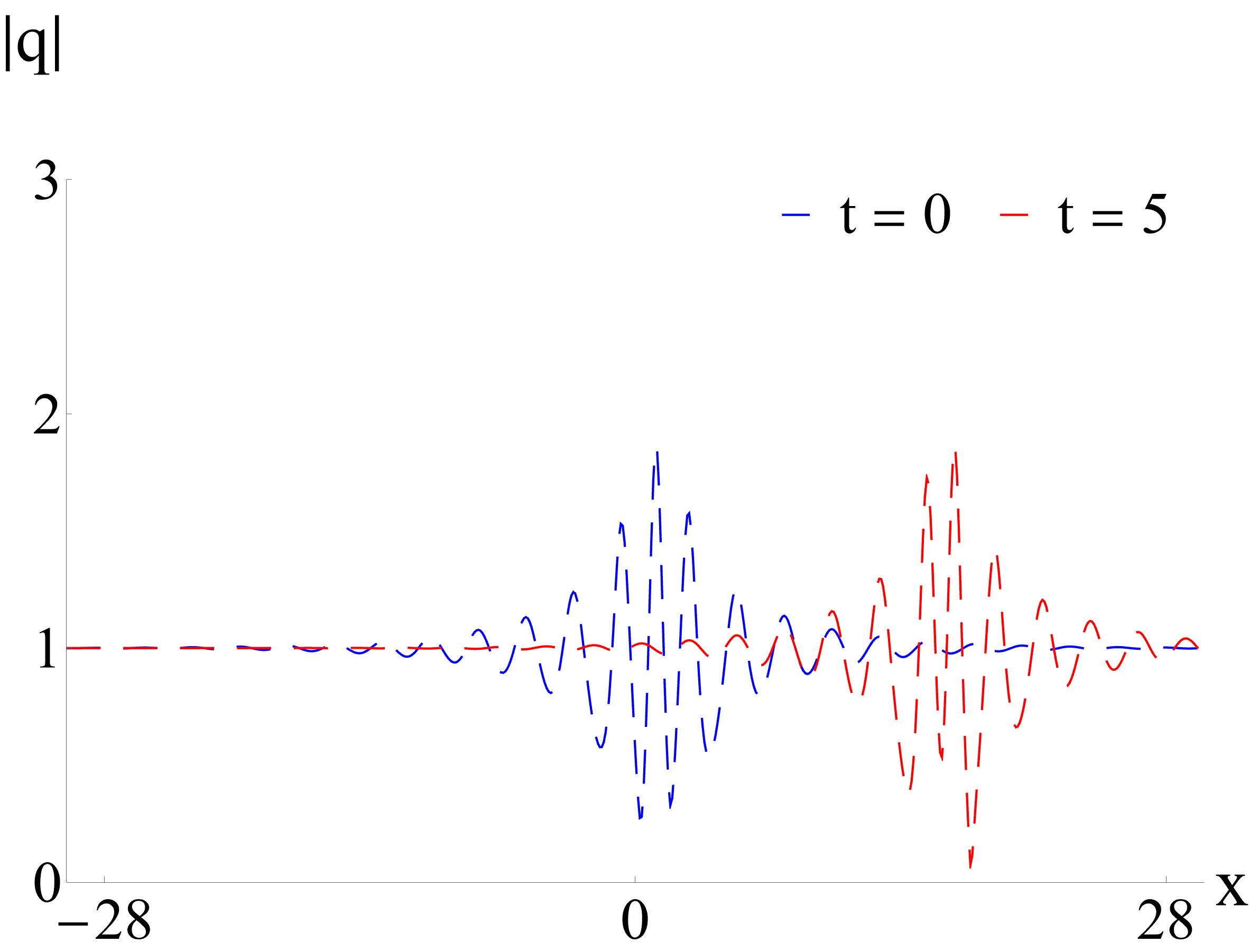}
		\end{minipage}
	\end{minipage}
	
	\vspace{0.5\baselineskip} 
	
	\noindent\begin{minipage}{\linewidth}
		\begin{minipage}[c]{0.01\linewidth}  
			\centering
			\footnotesize (b)
		\end{minipage}
		\hfill
		\begin{minipage}[c]{0.43\linewidth}  
			\centering
			\includegraphics[width=\linewidth]{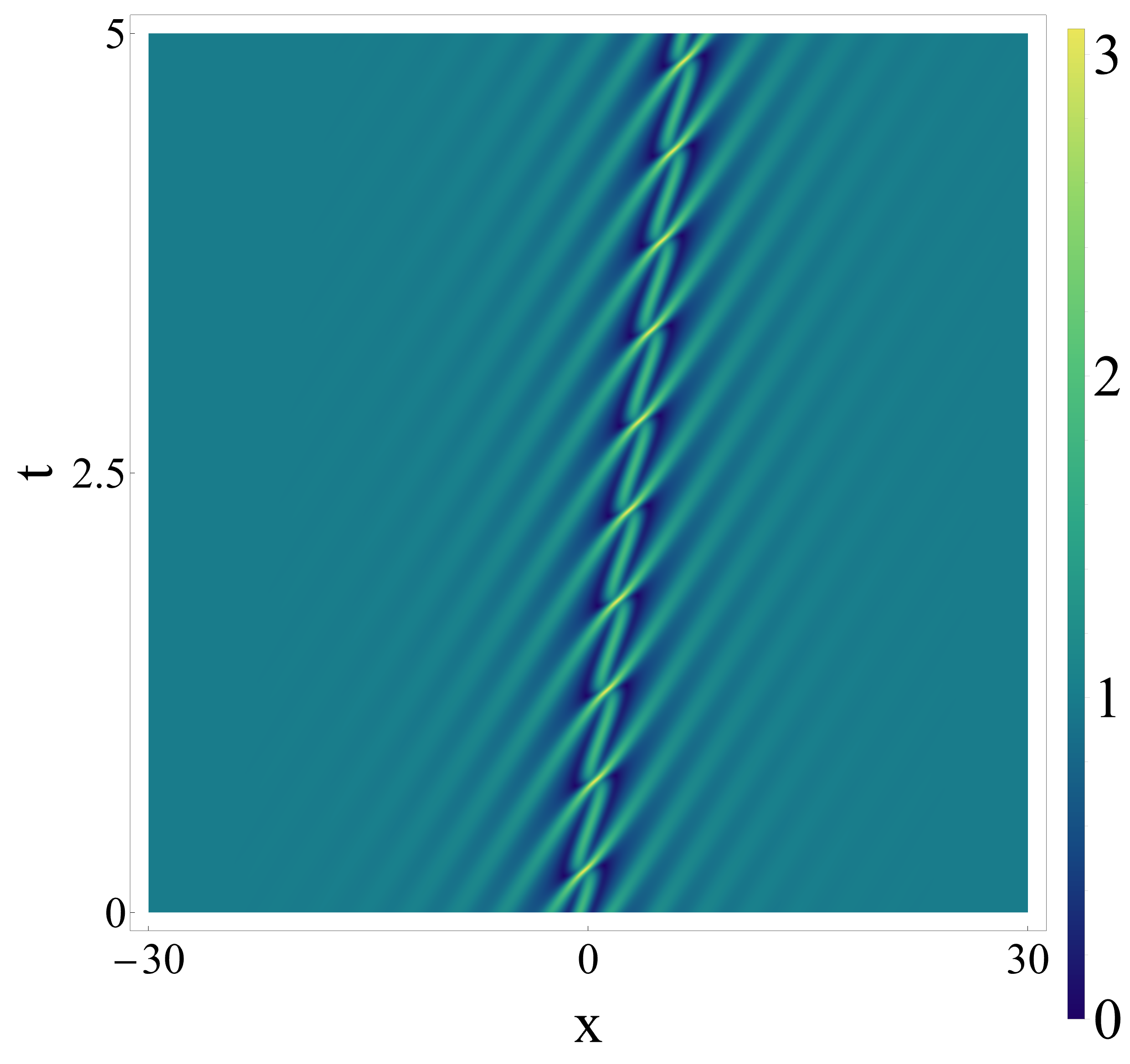}
		\end{minipage}
		\hfill
		\begin{minipage}[c]{0.01\linewidth}  
			\centering
			\footnotesize (d)
		\end{minipage}
		\hfill
		\begin{minipage}[c]{0.43\linewidth}  
			\centering
			\includegraphics[width=\linewidth]{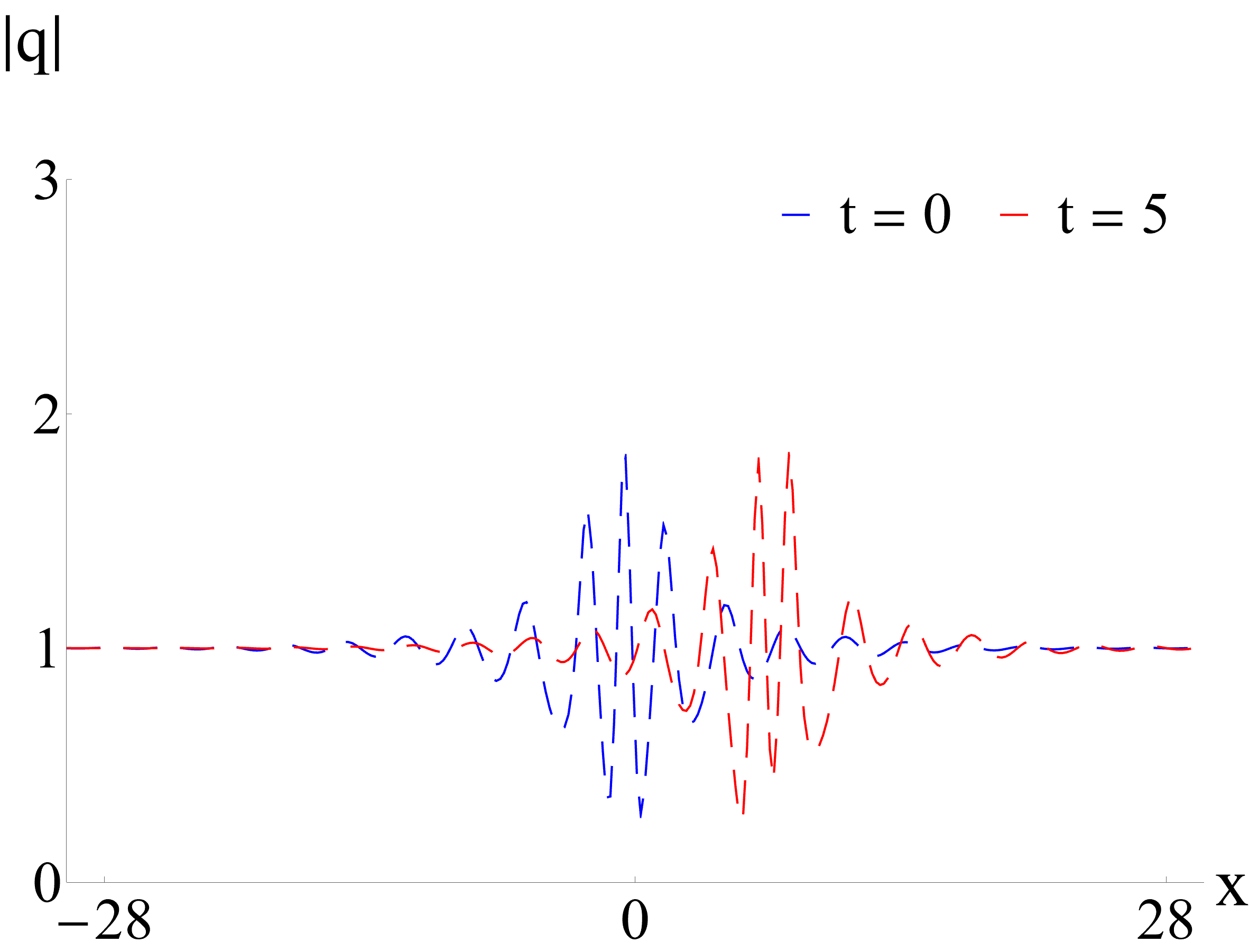}
		\end{minipage}
	\end{minipage}
	\caption{Density and amplitude profiles of quasi-ABs. Panels (a) and (c) correspond to the purple region with $b=1$ and $\omega=2.5$; panels (b) and (d) correspond to the green region with $b=0.4$ and $\omega=2$. Panels (a) and (b) display the density $|q(x, t; \chi_1)|$, while panels (c) and (d) show the corresponding amplitude profiles. Common parameters: $a=1$, $\sigma=1$, $\gamma=0.1$, $x_{01}=0$ and $t_{01}=0$.}
	\label{fig:zhunABs}
\end{figure}

The purple and green regions correspond to two distinct types of quasi-AB solutions emerging near the transition boundaries between different spectral regimes.
In Fig.~\ref{fig:zhunABs}, panels (a) and (b) display the density profiles of the solutions in the purple and green regions, respectively. Panels (c) and (d) show the corresponding amplitude profiles at $t=0$ (blue dashed lines) and $t=5$ (red dashed lines).

{Accordingly, in the present scalar reduction, second-order SRBs require the focusing, MI-active sector specified in Sec.~V.A, whereas the defocusing regime does not support an SRB.}
%


\section{Numerical Excitation of AB Solutions}
The spectral analysis in the previous section identifies the admissible parameter regions for different breather structures. To verify that these nonlinear excitations can be generated from small perturbations of the continuous-wave background, we perform direct numerical simulations using localized and periodic initial perturbations. Following previous studies, the initial excitation is imposed as
\begin{equation}
	\label{eq:initial}
	q=q_0\left(1+\epsilon \mathcal{L}_p(x/x_w)\cos(\omega x)\exp(\mathrm{i}\phi)\right),
\end{equation}
where $q_0$ denotes the continuous-wave background defined in Eq.~\eqref{eq:solution}, and the perturbation envelope $\mathcal L_p$ is chosen as either a hyperbolic-secant function, $\mathcal{L}_p = \sech(x/x_w)$, or a Gaussian, $\mathcal{L}_p = \exp(-x^2/x_w^2)$. The parameter $x_w$, which controls the spatial localization width of the perturbation, is set to $x_w \sim 1/(2\gamma)$ to match the envelope width of the analytical solutions. The carrier frequency $\omega$ is taken from the corresponding exact solution and $\epsilon \ll 1$ governs the perturbation amplitude. For the strictly periodic AB solution, we set $\mathcal{L}_p = 1$ in the numerical simulations.
\begin{figure}[htbp]
	\centering
	\noindent\begin{minipage}{\linewidth}
		\begin{minipage}[c]{0.01\linewidth}  
			\centering
			\footnotesize (a)
		\end{minipage}
		\hfill
		\begin{minipage}[c]{0.43\linewidth} 
			\centering
			\includegraphics[width=\linewidth]{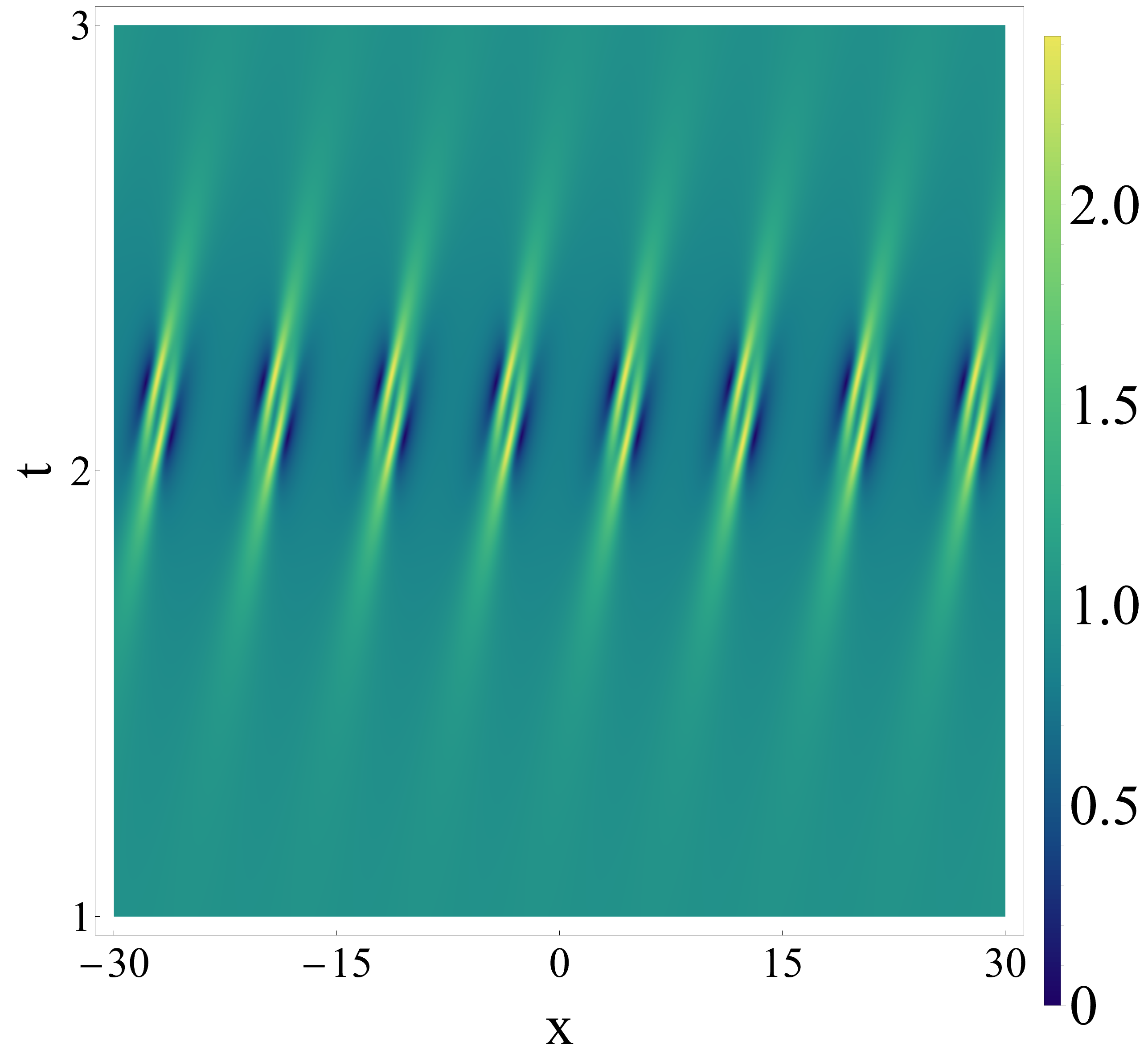}
		\end{minipage}
		\hfill
		\begin{minipage}[c]{0.01\linewidth}  
			\centering
			\footnotesize (c)
		\end{minipage}
		\hfill
		\begin{minipage}[c]{0.43\linewidth}  
			\centering
			\includegraphics[width=\linewidth]{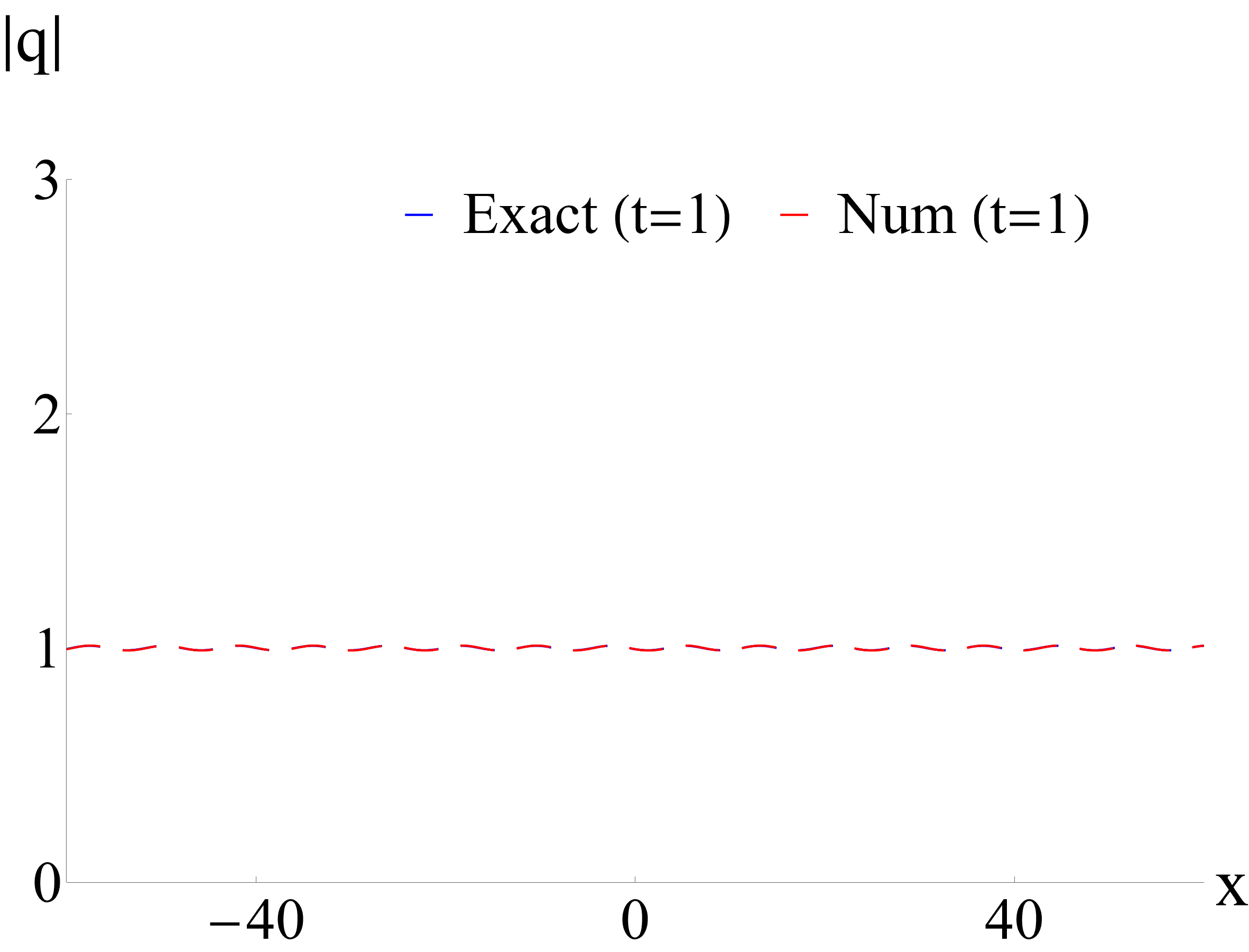}
		\end{minipage}
	\end{minipage}
	
	\vspace{0.5\baselineskip} 
	
	\noindent\begin{minipage}{\linewidth}
		\begin{minipage}[c]{0.01\linewidth}  
			\centering
			\footnotesize (b)
		\end{minipage}
		\hfill
		\begin{minipage}[c]{0.43\linewidth}  
			\centering
			\includegraphics[width=\linewidth]{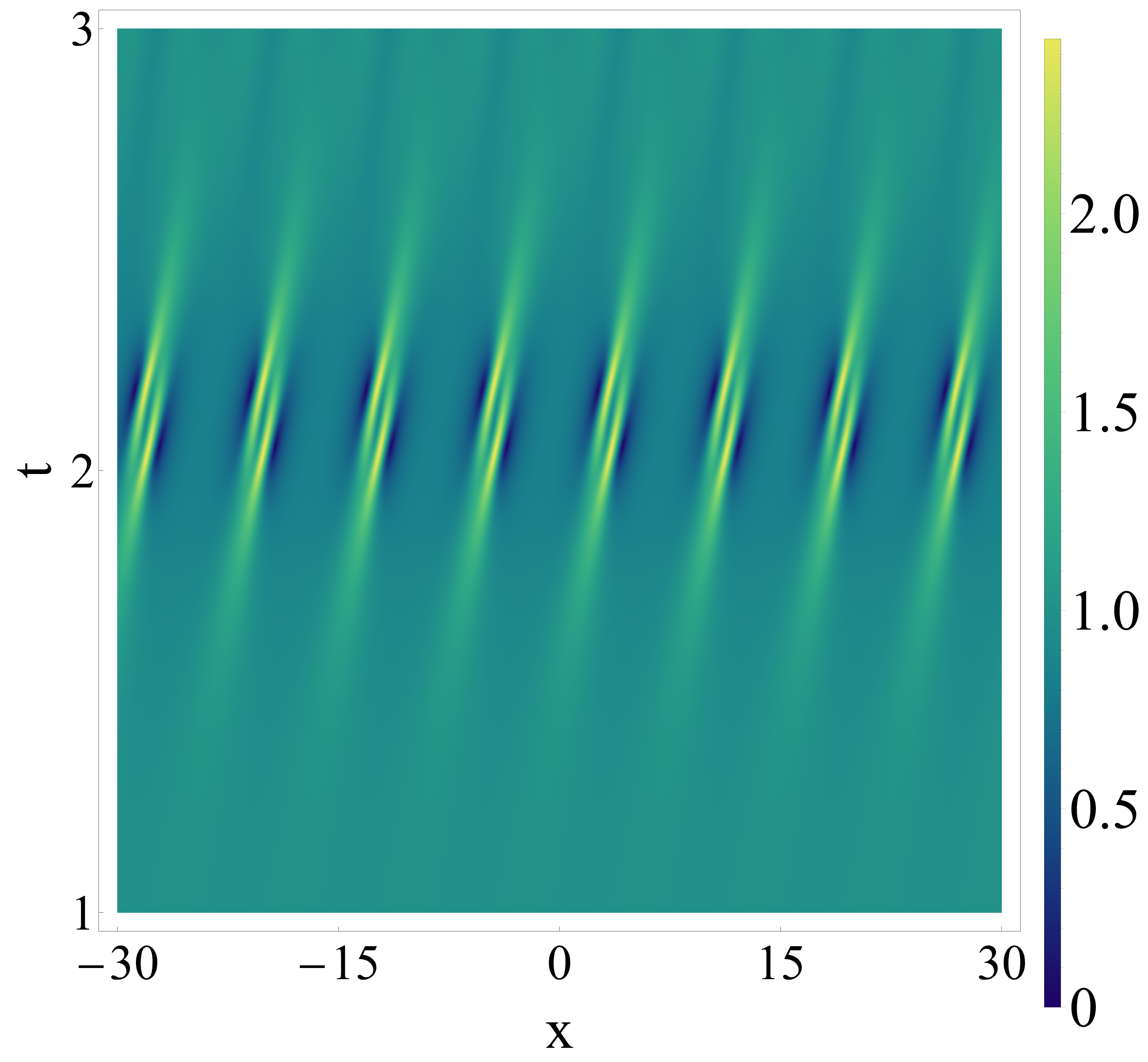}
		\end{minipage}
		\hfill
		\begin{minipage}[c]{0.01\linewidth}  
			\centering
			\footnotesize (d)
		\end{minipage}
		\hfill
		\begin{minipage}[c]{0.43\linewidth}  
			\centering
			\includegraphics[width=\linewidth]{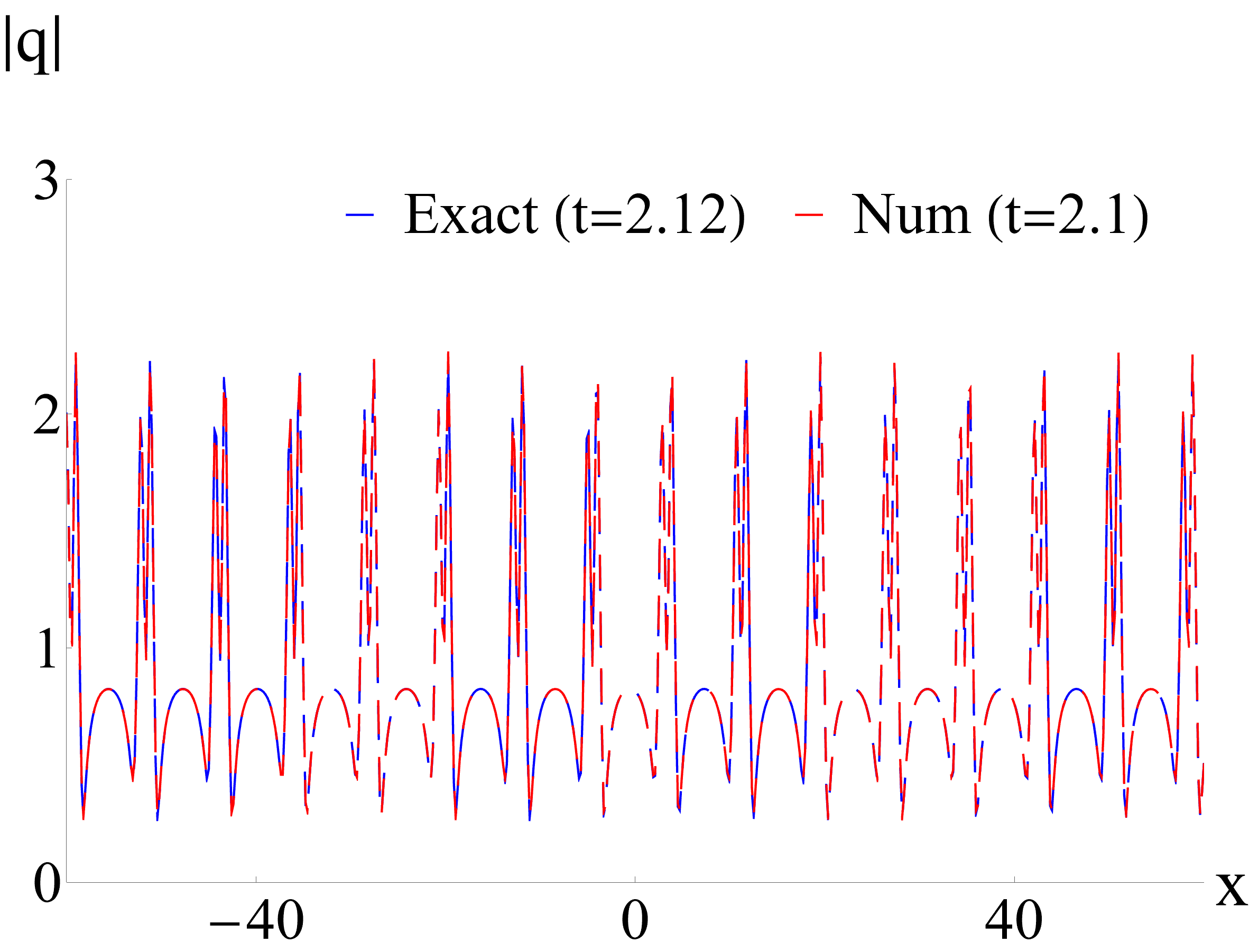}
		\end{minipage}
	\end{minipage}
	\caption{Exact and numerical density and amplitude maps of four-petal ABs. Panels (a) and (b) show the exact and numerical density $|q(x, t; \chi_1)|$ of four-petaled ABs, respectively; panels (c) and (d) display the corresponding amplitude profiles. Parameters: $a=1$, $b=1$, $\sigma=1$, $\gamma=0$, $\omega=0.8$, $\phi=3 \pi/2$, $\epsilon=0.001$, $x_{01}=-1$ and $t_{01}=2.2$.}
	\label{fig:ABs1}
\end{figure}

The four-petal AB solution corresponds to the blue region in Fig. \ref{fig:myLabel}. As shown in Fig. \ref{fig:ABs1}, with $\mathcal{L}_p = 1$ and the strict periodicity of the AB solution, the initial conditions excite a solution nearly identical to the exact one, apart from translational shifts in space and time. Panel (a) displays the density of the exact solution and panel (b) shows the corresponding numerical density. Panels (c) and (d) compare the amplitude profiles: at $t=1$, the exact (blue dashed) and numerical (red dashed) solutions are shown in (c); at $t=2.12$ (exact) and $t=2.1$ (numerical), the comparison is shown in (d).

\begin{figure}[htbp]
	\centering
	\noindent\begin{minipage}{\linewidth}
		\begin{minipage}[c]{0.01\linewidth}  
			\centering
			\footnotesize (a)
		\end{minipage}
		\hfill
		\begin{minipage}[c]{0.43\linewidth} 
			\centering
			\includegraphics[width=\linewidth]{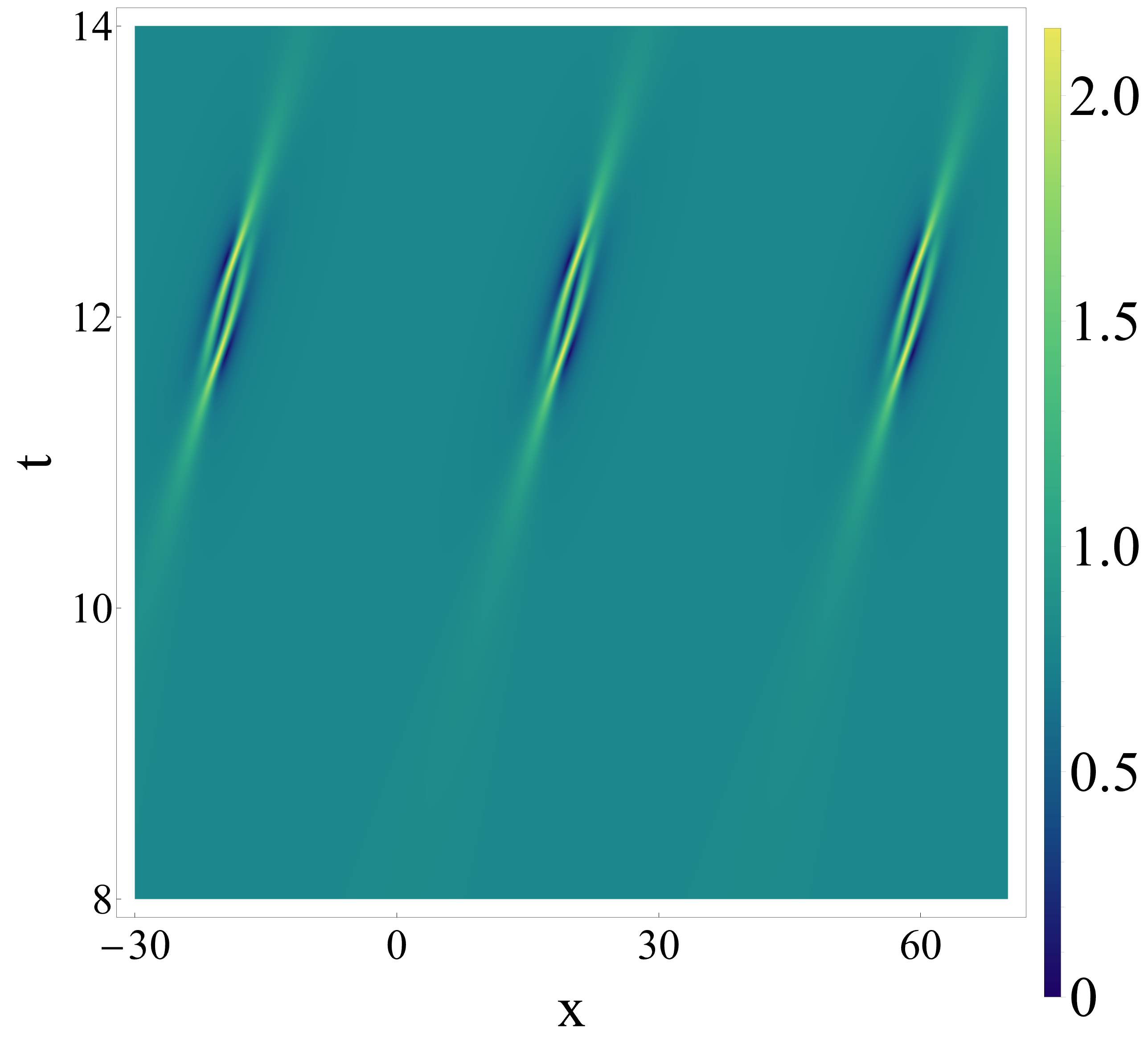}
		\end{minipage}
		\hfill
		\begin{minipage}[c]{0.01\linewidth}  
			\centering
			\footnotesize (c)
		\end{minipage}
		\hfill
		\begin{minipage}[c]{0.43\linewidth}  
			\centering
			\includegraphics[width=\linewidth]{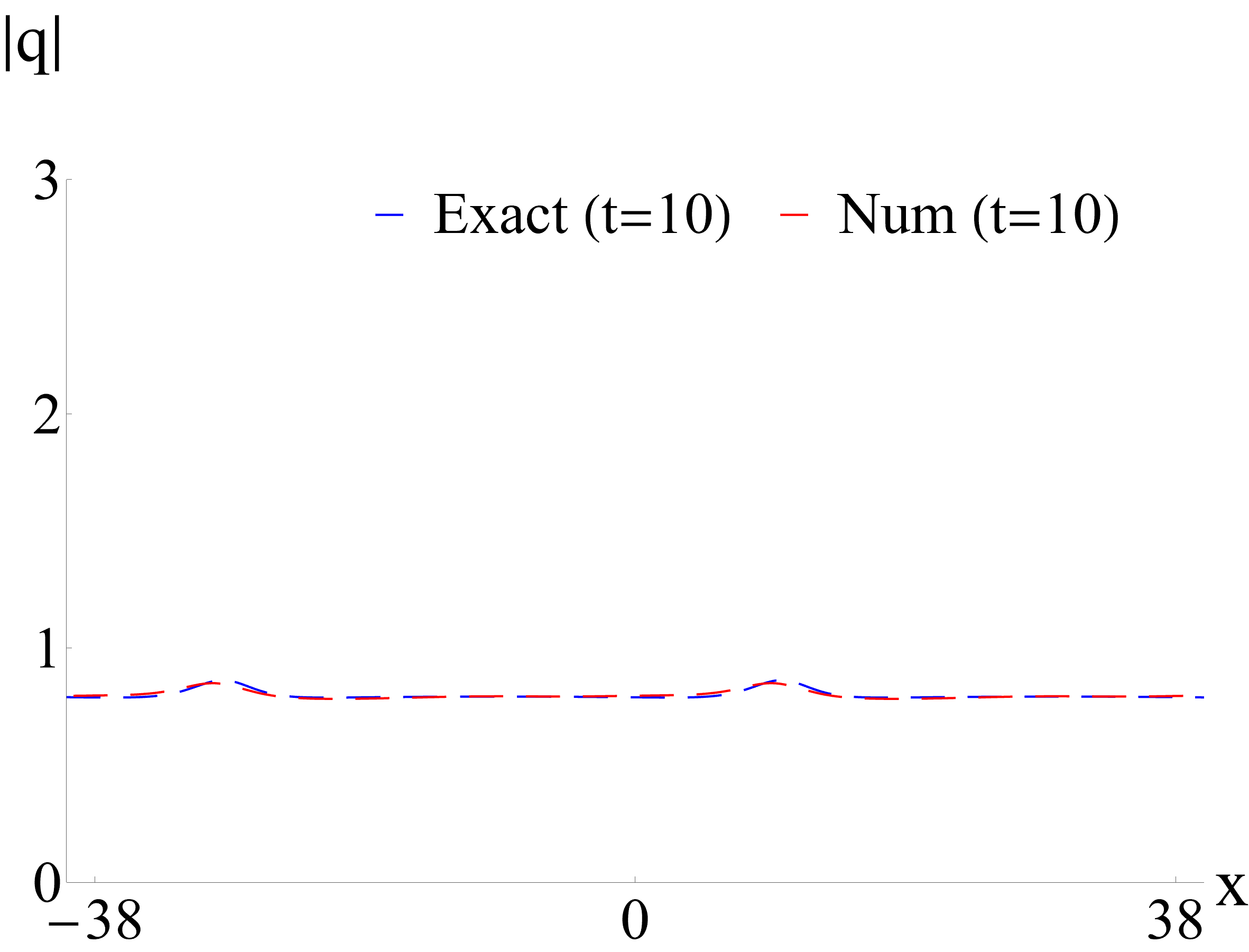}
		\end{minipage}
	\end{minipage}
	
	\vspace{0.5\baselineskip} 
	
	\noindent\begin{minipage}{\linewidth}
		\begin{minipage}[c]{0.01\linewidth}  
			\centering
			\footnotesize (b)
		\end{minipage}
		\hfill
		\begin{minipage}[c]{0.43\linewidth}  
			\centering
			\includegraphics[width=\linewidth]{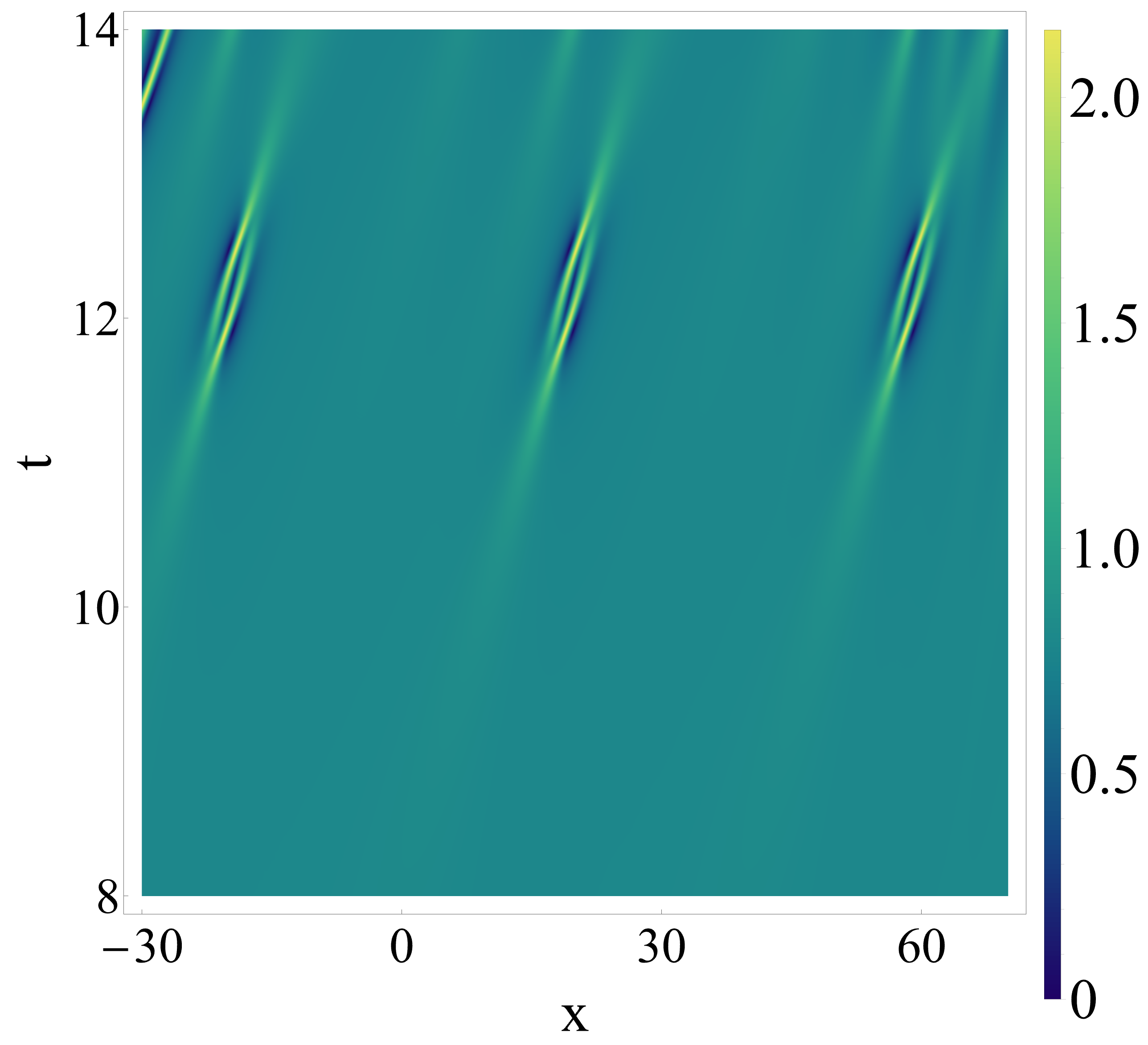}
		\end{minipage}
		\hfill
		\begin{minipage}[c]{0.01\linewidth}  
			\centering
			\footnotesize (d)
		\end{minipage}
		\hfill
		\begin{minipage}[c]{0.43\linewidth}  
			\centering
			\includegraphics[width=\linewidth]{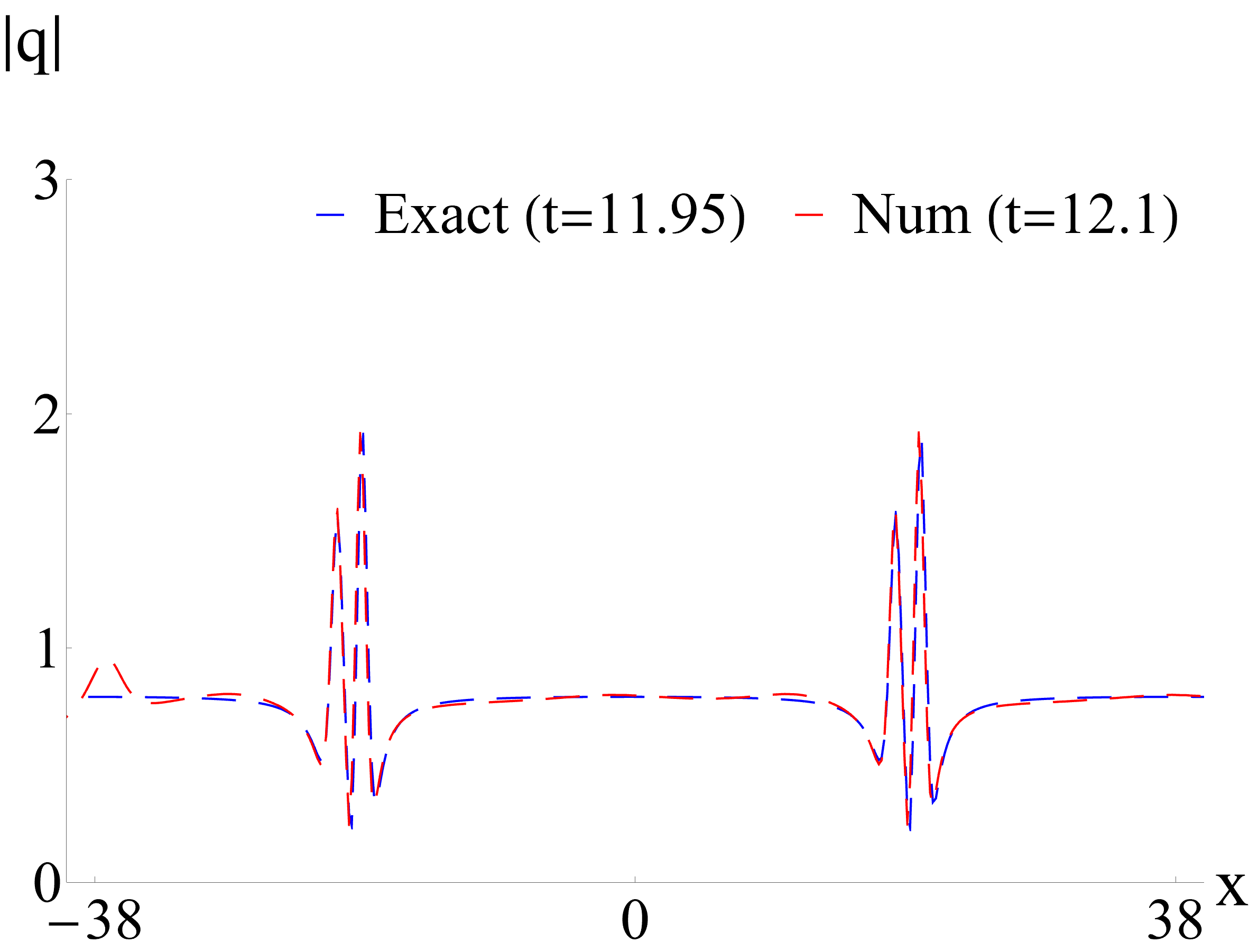}
		\end{minipage}
	\end{minipage}
	\caption{Exact and numerical density and amplitude maps of twisted ABs. Panels (a) and (b) show the exact and numerical density $|q(x, t; \chi_1)|$ of twisted ABs, respectively; panels (c) and (d) display the corresponding amplitude profiles. Parameters: $a=0.8$, $b=0.6$, $\sigma=1$, $\gamma=0$, $\omega=0.16$, $\phi=3 \pi/2$, $\epsilon=0.01$, $x_{01}=0.5$ and $t_{01}=12.4$.}
	\label{fig:ABs2}
\end{figure}
Similarly, with the periodic perturbation envelope $\mathcal L_p=1$, a twisted AB solution that closely matches the exact solution is excited in Fig. \ref{fig:ABs2}. Panels (a) and (b) display the density of the exact and numerical solutions, respectively. Panels (c) and (d) compare the corresponding amplitude profiles: at $t=10$ in (c) and at $t=11.95$ (exact) versus $t=12.1$ (numerical) in (d); blue and red dashed lines denote the exact and numerical results, respectively. {Over the displayed interval, the numerical profile follows the corresponding exact solution.}

\section{Spectral Mechanism and Numerical Excitation of SRB Solutions}
Since SRB solutions arise from the nonlinear superposition of multiple quasi-AB solutions, they are spatially localized rather than strictly periodic. Accordingly, their numerical excitation is implemented using a localized perturbation envelope $\mathcal{L}_p = \sech(x/x_w)$ or $\mathcal{L}_p = \exp(-x^2/x_w^2)$.
\subsection{Spectral condition for SRB formation}
{For $\sigma=1$ and $\gamma\to0$, it is useful to write}
{
\begin{align}
 n_0\equiv n\big|_{\gamma=0}&=a^4-b^2(4a^2-\omega^2),\\
 m_0\equiv m\big|_{\gamma=0}&=\frac{\omega^2}{4}+b^2-a^2 .
\end{align}
}
{The case $b=0$ is therefore only a special endpoint and cannot represent the full $n_0\geq0$ sector. For $b\neq0$ and $n_0\geq0$, define $r_\pm=m_0\pm\sqrt{n_0}$. The roots $\chi_{1,2}$ have nonzero imaginary parts only when $r_+<0$, while $\chi_{3,4}$ have nonzero imaginary parts only when $r_-<0$. Hence $r_-\geq0$ gives no quasi-AB pair, $r_+\geq0>r_-$ gives only the $\chi_{3,4}$ pair, and $r_+<0$ gives two spectral quasi-AB pairs. Equality corresponds to a coalescent, marginal branch. This completes the root classification for the focusing $n_0\geq0$, $b\neq0$ sector; the MI calculation below shows that its linear gain vanishes, so these spectral candidates are not identified as SRBs by the present excitation criterion.}

{For $n_0<0$, $\chi_i^+=-\chi_i^-$ and $\rho^+=\rho^-$. From the relation between $\lambda$ and $\chi$, it follows that $\lambda_{1r}=\lambda_{3r}$, $\lambda_{2r}=\lambda_{4r}$, $\lambda_{1i}=-\lambda_{3i}$, and $\lambda_{2i}=-\lambda_{4i}$. Under these symmetry constraints, all four eigenvalues $\chi_1,\chi_2,\chi_3,\chi_4$ are admissible for the paired quasi-AB construction that produces the second-order SRB.}

\subsection{Relation between MI growth rate and group velocity difference}
To establish the relation between MI growth and SRB formation, we derive the instability growth rate from the group velocity expression. The exponential amplification factor of the quasi-AB envelope is determined by the imaginary part of the phase variable $\tau$, yielding
\begin{equation}
	\label{eq:MI}
	\mathcal{G}=\left\lvert 2\gamma V_g(\chi)\right\rvert=\left\lvert \omega\chi' \right\rvert ,\quad
	\chi'\equiv 4\lambda_r\chi_i+2\lambda_i(2\chi_r+\omega) |_{\gamma=0}.
\end{equation}
Meanwhile, from Eq.\eqref{eq:chiri}, we have
\begin{equation}
	\Delta V_g(\chi_1;\chi_3)=\Delta V_g(\chi_2;\chi_4)=\left\lvert \frac{\omega}{\gamma} \chi_i^- \right\rvert,
\end{equation}
therefore,the growth rate of MI is
\begin{equation}
	\mathcal{G}=\gamma \Delta V_g(\chi_1;\chi_3)=\gamma \Delta V_g(\chi_2;\chi_4).
\end{equation}
Since the growth rates of the two pairs $(\chi_1,\chi_3)$ and $(\chi_2,\chi_4)$ are equal, both pairs are admissible for constructing SRB solutions. Their superposition yields a second-order SRB solution.
This result establishes a direct quantitative connection between the MI growth rate and the relative propagation velocity of the two constituent quasi-ABs.

{To verify this result independently of the DT group-velocity calculation, we linearize Eq.~\eqref{eq:eq} about $q_0=a\exp(\mathrm{i}\theta_1)$ and use a normal mode proportional to $\exp(\mathrm{i}\omega x+s t)$. Keeping the two coupled sidebands gives}
{
\begin{equation}
\left[s+\mathrm{i}\omega\left(9\sigma a^2-3b^2-\omega^2\right)\right]^2=-9\omega^2 n_0 .
\label{eq:linearMI}
\end{equation}
}
{Thus the independently obtained linear MI gain is $\mathcal{G}_{\rm lin}=3|\omega|\sqrt{-n_0}$ for $n_0<0$ and is zero for $n_0\geq0$. Substitution of the roots in Eq.~\eqref{eq:eigenvalue} into the definition in Eq.~\eqref{eq:MI} gives $|\chi'|=3\sqrt{-n_0}$ in the MI-active sector. Hence $\mathcal{G}_{\rm lin}=|\omega\chi'|=\gamma\Delta V_g$, establishing the equality between the linear MI gain and the group-velocity expression.}

{For comparison, the formal $\epsilon\to\infty$ limit of Eq.~\eqref{eq:eq0} is the focusing NLS equation. Its standard plane-wave linearization gives $\mathcal{G}_{\rm NLS}(\kappa)=\frac{|\kappa|}{2}\sqrt{4a^2-\kappa^2}$ for $0<|\kappa|<2a$. The SS condition $n_0<0$ depends explicitly on $b$ and is therefore not the NLS condition: the higher-order SS terms together with the background carrier wavenumber reshape the MI-active spectral region.}
\subsection{Numerical Excitation}
\begin{figure}[htbp]
	\centering
	\noindent\begin{minipage}{\linewidth}
		\begin{minipage}[c]{0.01\linewidth}  
			\centering
			\footnotesize (a)
		\end{minipage}
		\hfill
		\begin{minipage}[c]{0.43\linewidth} 
			\centering
			\includegraphics[width=\linewidth]{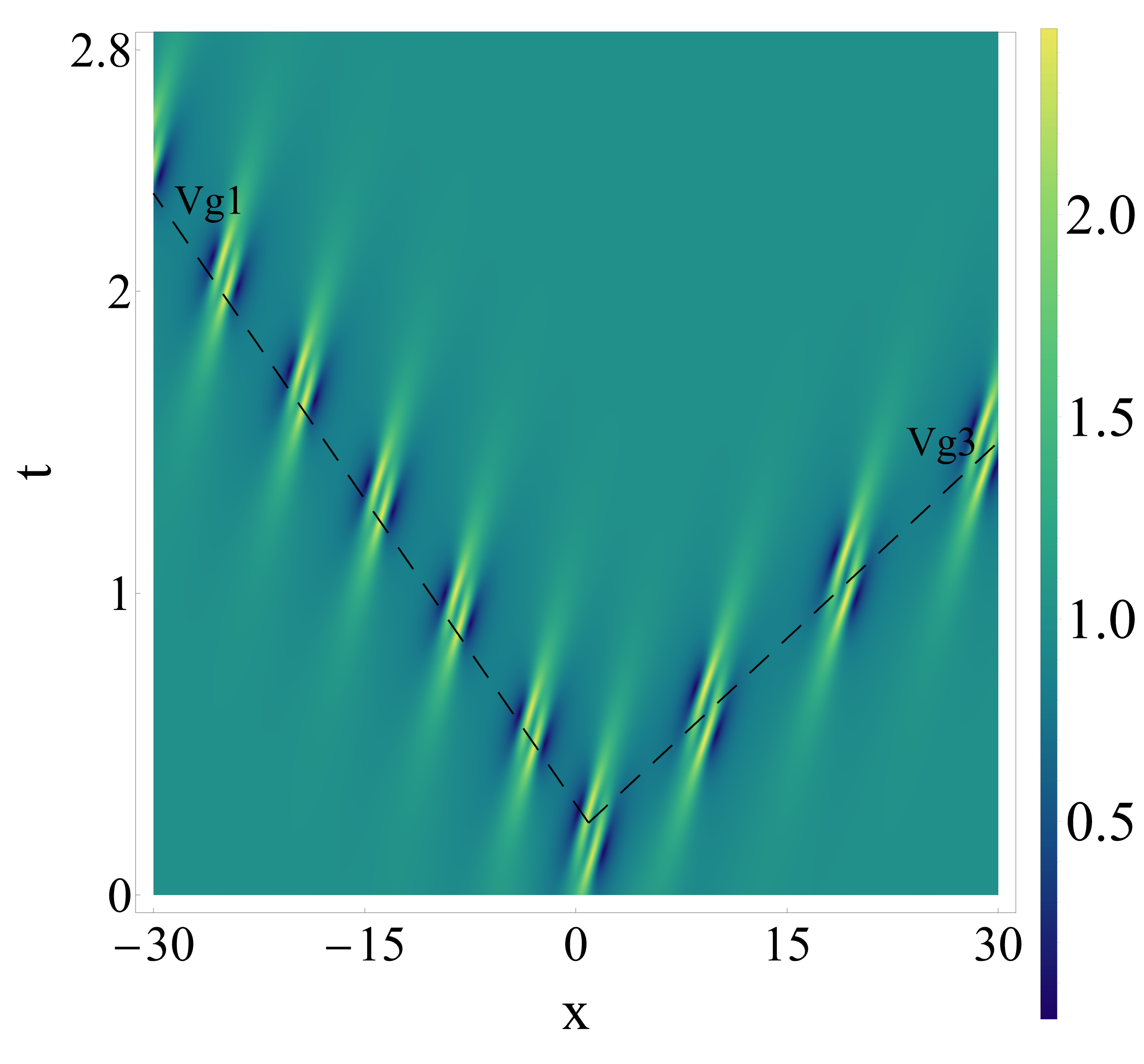}
		\end{minipage}
		\hfill
		\begin{minipage}[c]{0.01\linewidth}  
			\centering
			\footnotesize (c)
		\end{minipage}
		\hfill
		\begin{minipage}[c]{0.43\linewidth}  
			\centering
			\includegraphics[width=\linewidth]{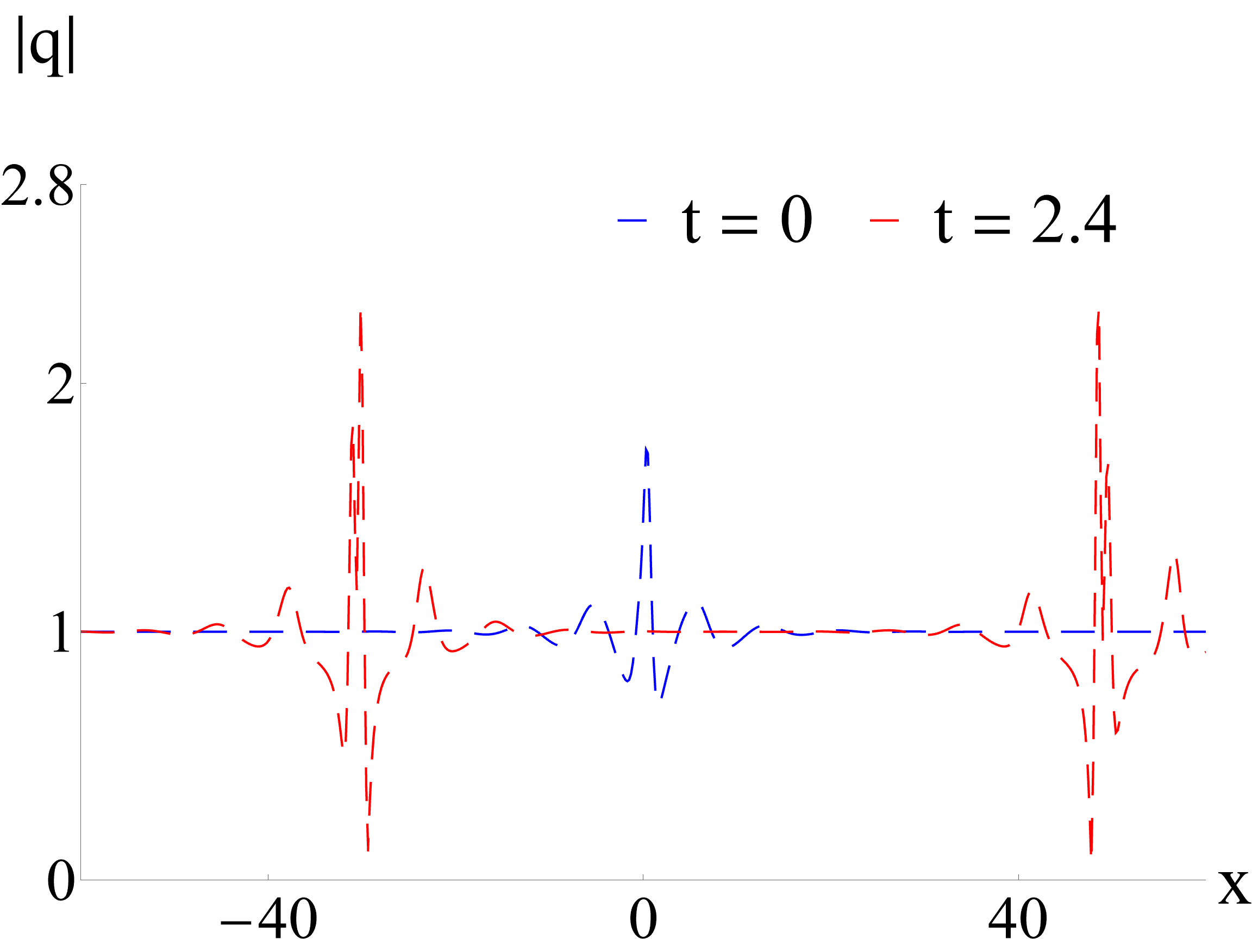}
		\end{minipage}
	\end{minipage}
	
	\vspace{0.5\baselineskip} 
	
	\noindent\begin{minipage}{\linewidth}
		\begin{minipage}[c]{0.01\linewidth}  
			\centering
			\footnotesize (b)
		\end{minipage}
		\hfill
		\begin{minipage}[c]{0.43\linewidth}  
			\centering
			\includegraphics[width=\linewidth]{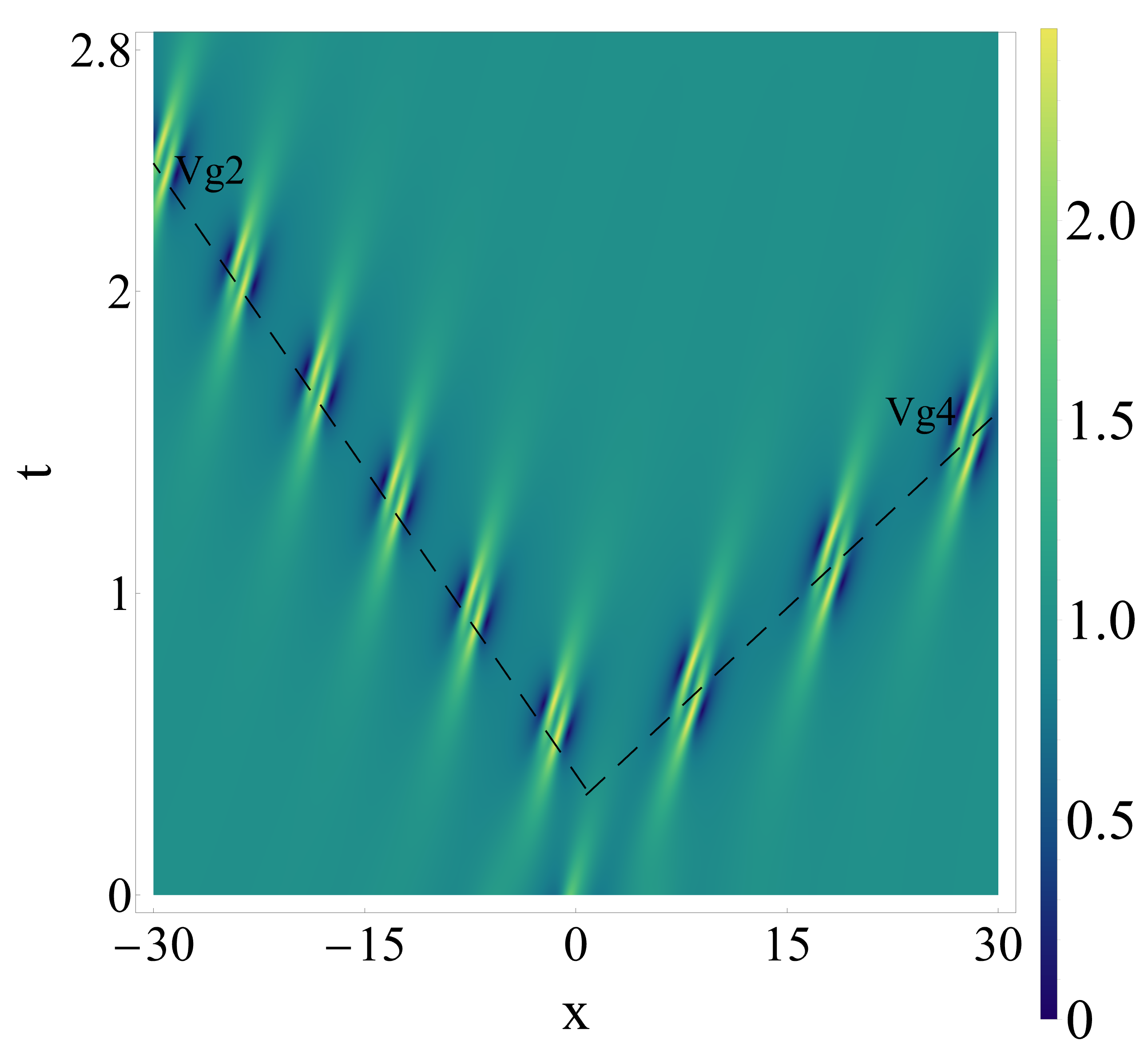}
		\end{minipage}
		\hfill
		\begin{minipage}[c]{0.01\linewidth}  
			\centering
			\footnotesize (d)
		\end{minipage}
		\hfill
		\begin{minipage}[c]{0.43\linewidth}  
			\centering
			\includegraphics[width=\linewidth]{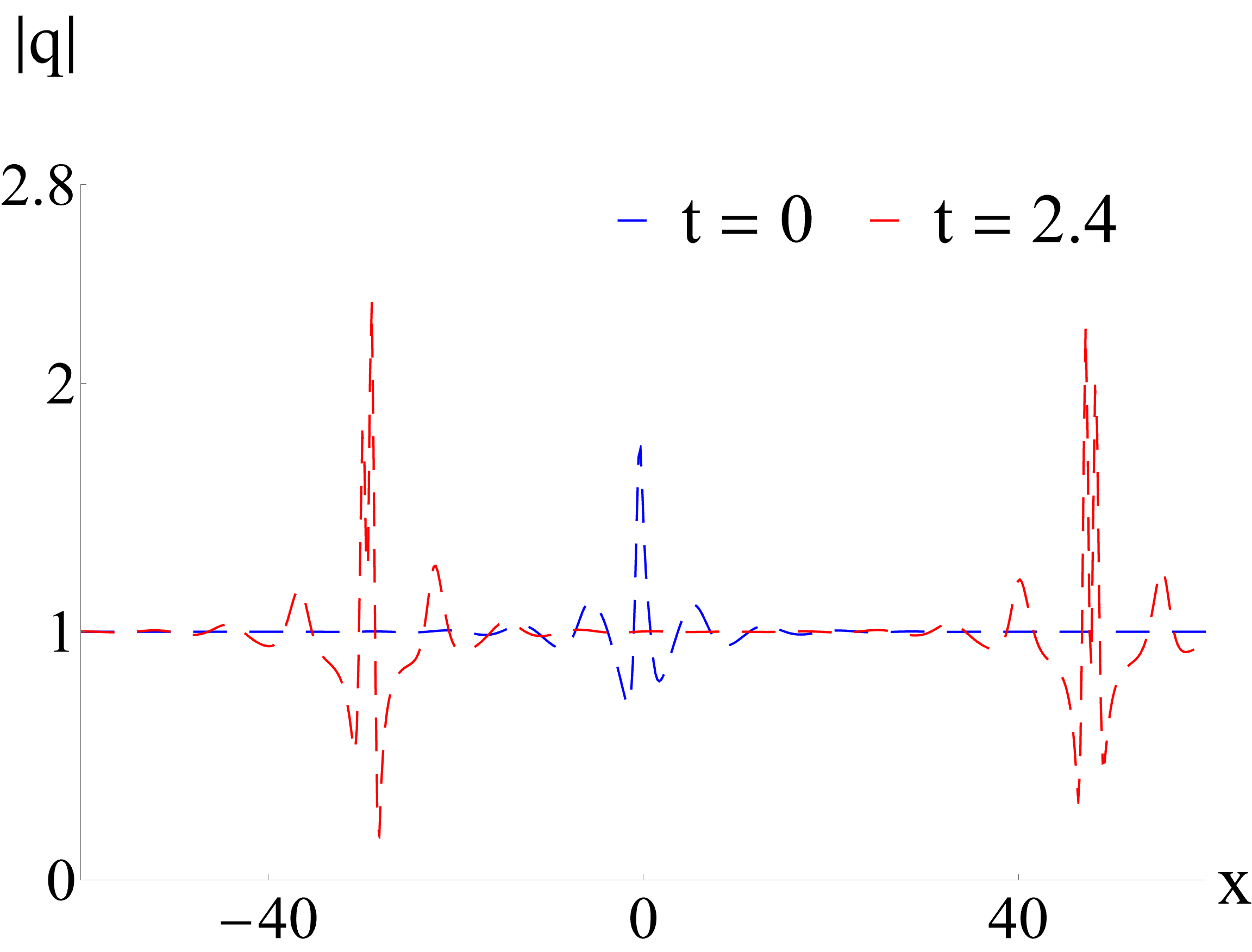}
		\end{minipage}
	\end{minipage}
	\caption{Density and amplitude profiles of the exact four-petal SRB solution. Panels (a) and (b) correspond to the density of $|q(x, t; \chi_1,\chi_3)|$ and $|q(x, t; \chi_2,\chi_4)|$, while panels (c) and (d) correspond to their respective amplitudes. The parameters are set to $a=1$, $b=1$, $\sigma=1$, $\gamma=0.1$, $\omega=0.8$ and $x_{01}=x_{02}=t_{01}=t_{02}=0$.}
	\label{fig:jingquejie}
\end{figure}

As shown in Fig. \ref{fig:jingquejie}, both solution branches exhibit a four-petal structure, with the amplitude distributions of the two branches showing noticeable differences. The labels $V_{g1}$--$V_{g4}$ mark the corresponding group velocities $V_g(\chi_1)$ through $V_g(\chi_4)$, respectively.

\begin{figure}[htbp]
	\centering
	\noindent\begin{minipage}{\linewidth}
		\begin{minipage}[c]{0.01\linewidth}  
			\centering
			\footnotesize (a)
		\end{minipage}
		\hfill
		\begin{minipage}[c]{0.43\linewidth} 
			\centering
			\includegraphics[width=\linewidth]{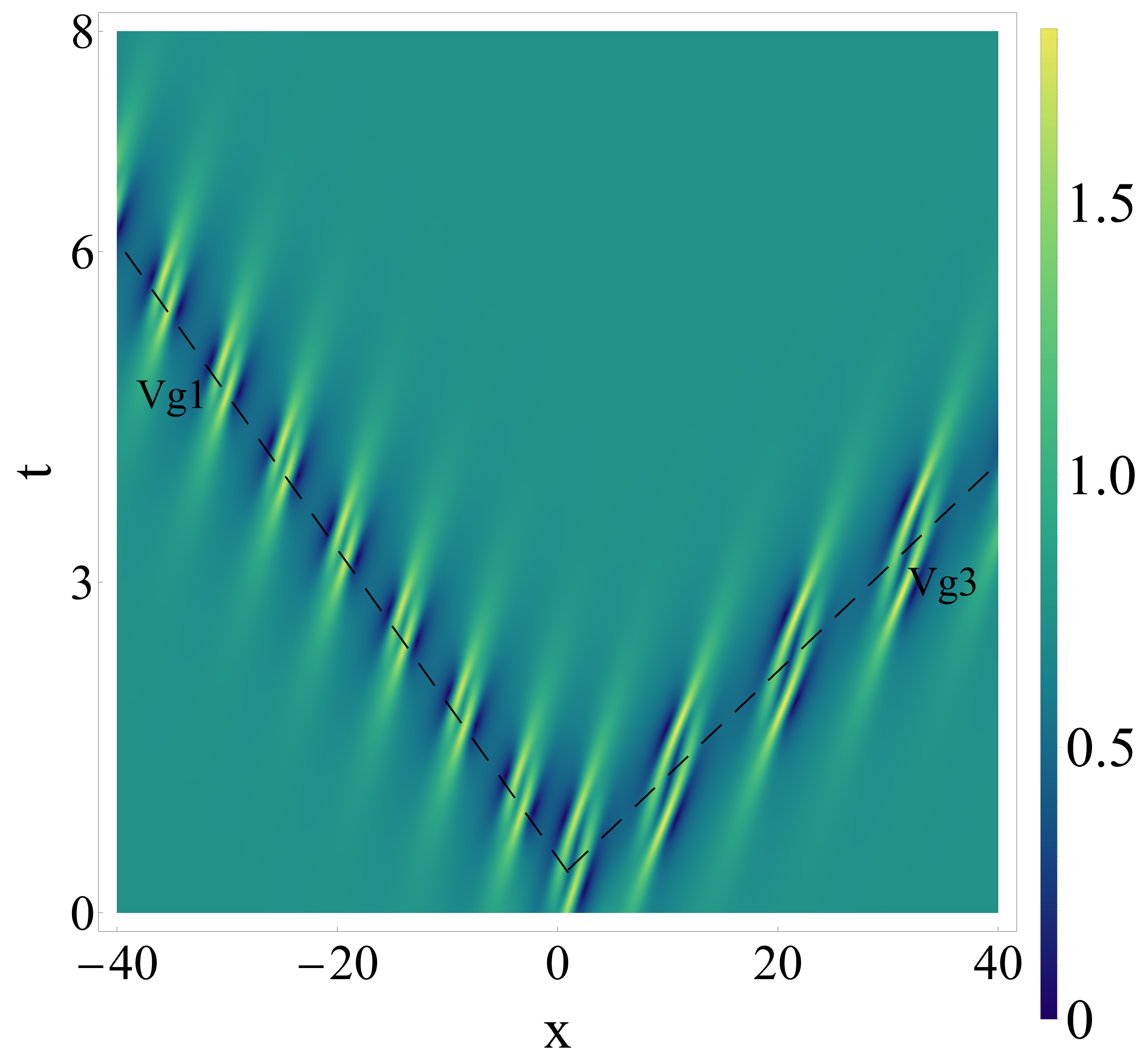}
		\end{minipage}
		\hfill
		\begin{minipage}[c]{0.01\linewidth}  
			\centering
			\footnotesize (c)
		\end{minipage}
		\hfill
		\begin{minipage}[c]{0.43\linewidth}  
			\centering
			\includegraphics[width=\linewidth]{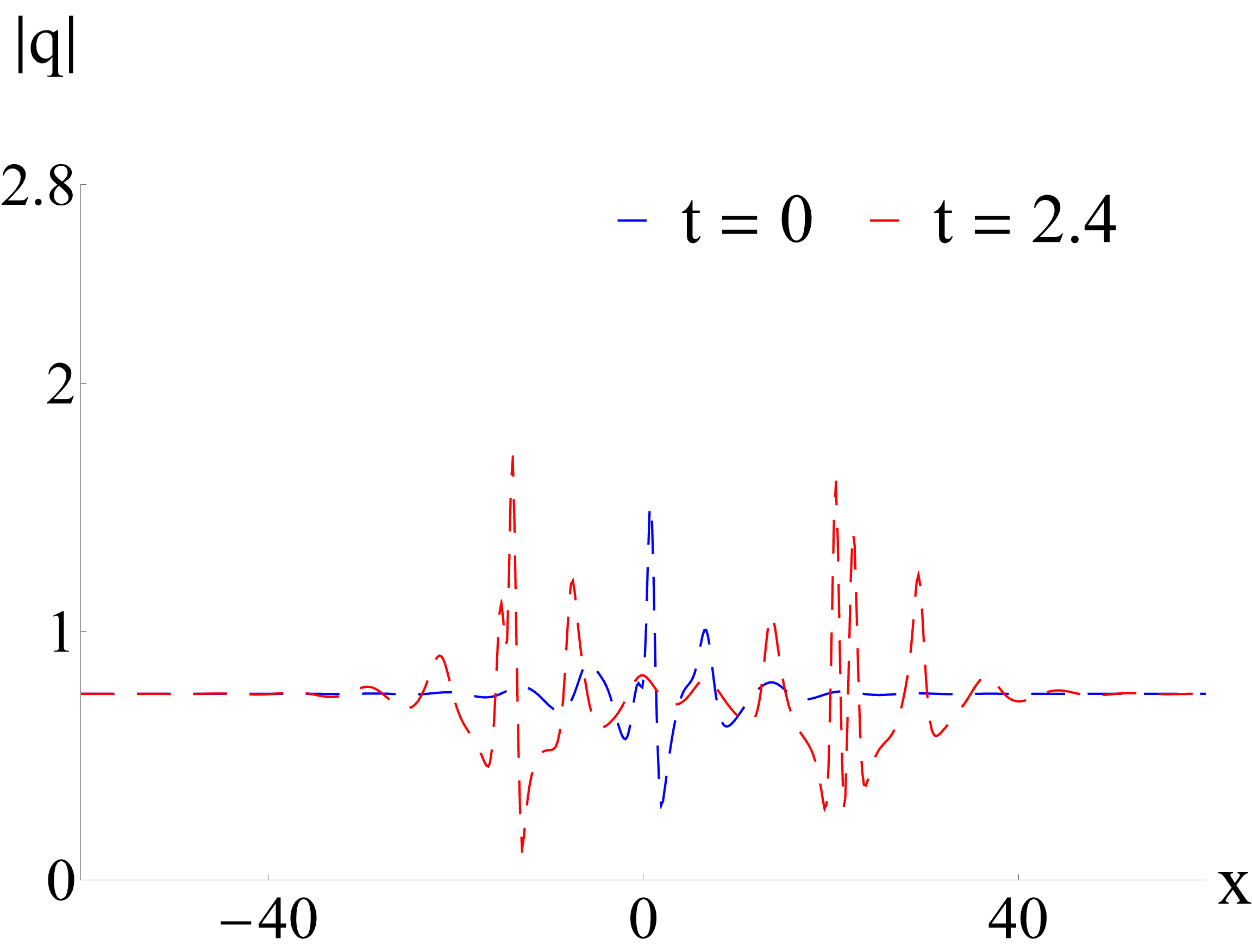}
		\end{minipage}
	\end{minipage}
	
	\vspace{0.5\baselineskip} 
	
	\noindent\begin{minipage}{\linewidth}
		\begin{minipage}[c]{0.01\linewidth}  
			\centering
			\footnotesize (b)
		\end{minipage}
		\hfill
		\begin{minipage}[c]{0.43\linewidth}  
			\centering
			\includegraphics[width=\linewidth]{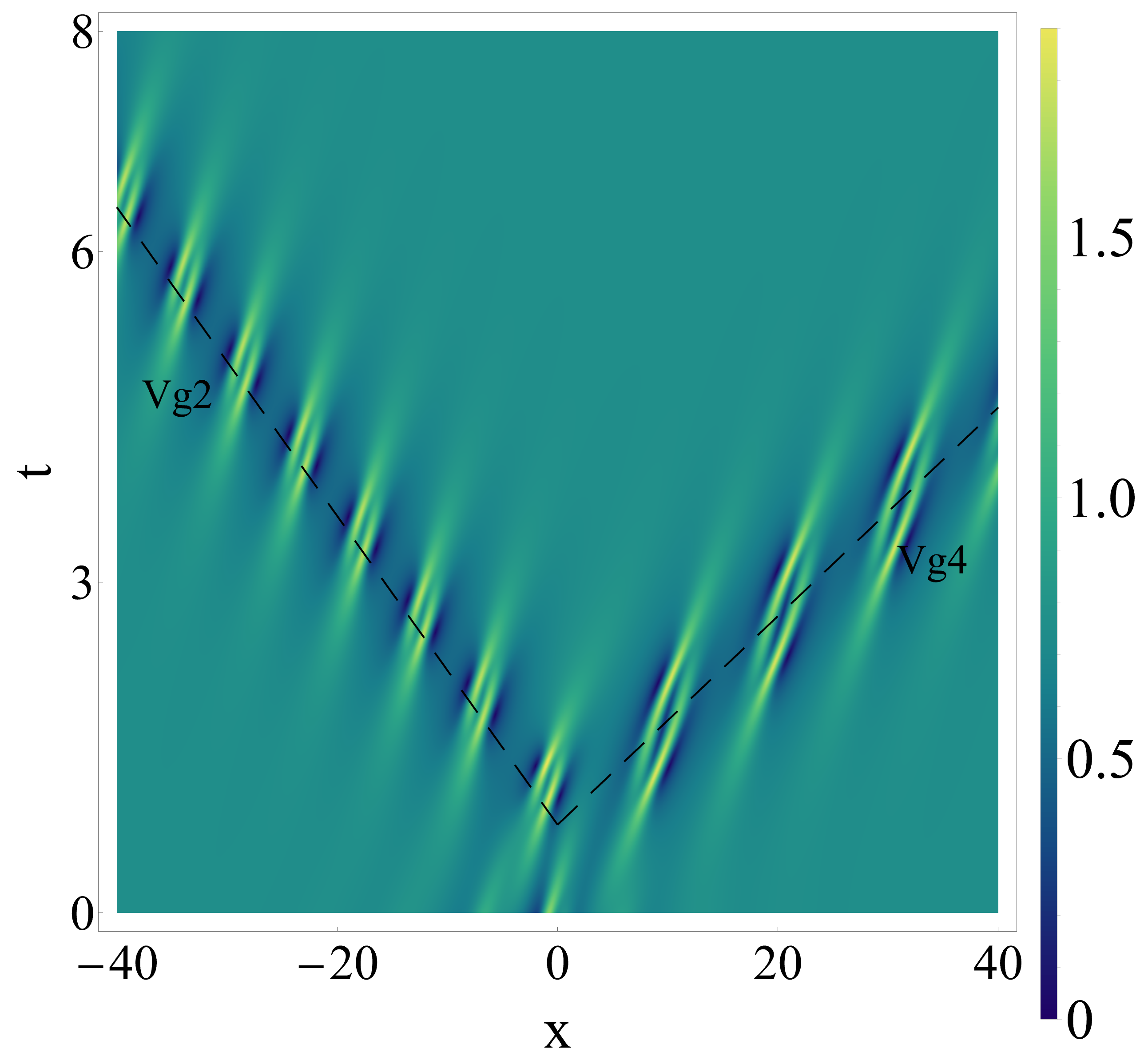}
		\end{minipage}
		\hfill
		\begin{minipage}[c]{0.01\linewidth}  
			\centering
			\footnotesize (d)
		\end{minipage}
		\hfill
		\begin{minipage}[c]{0.43\linewidth}  
			\centering
			\includegraphics[width=\linewidth]{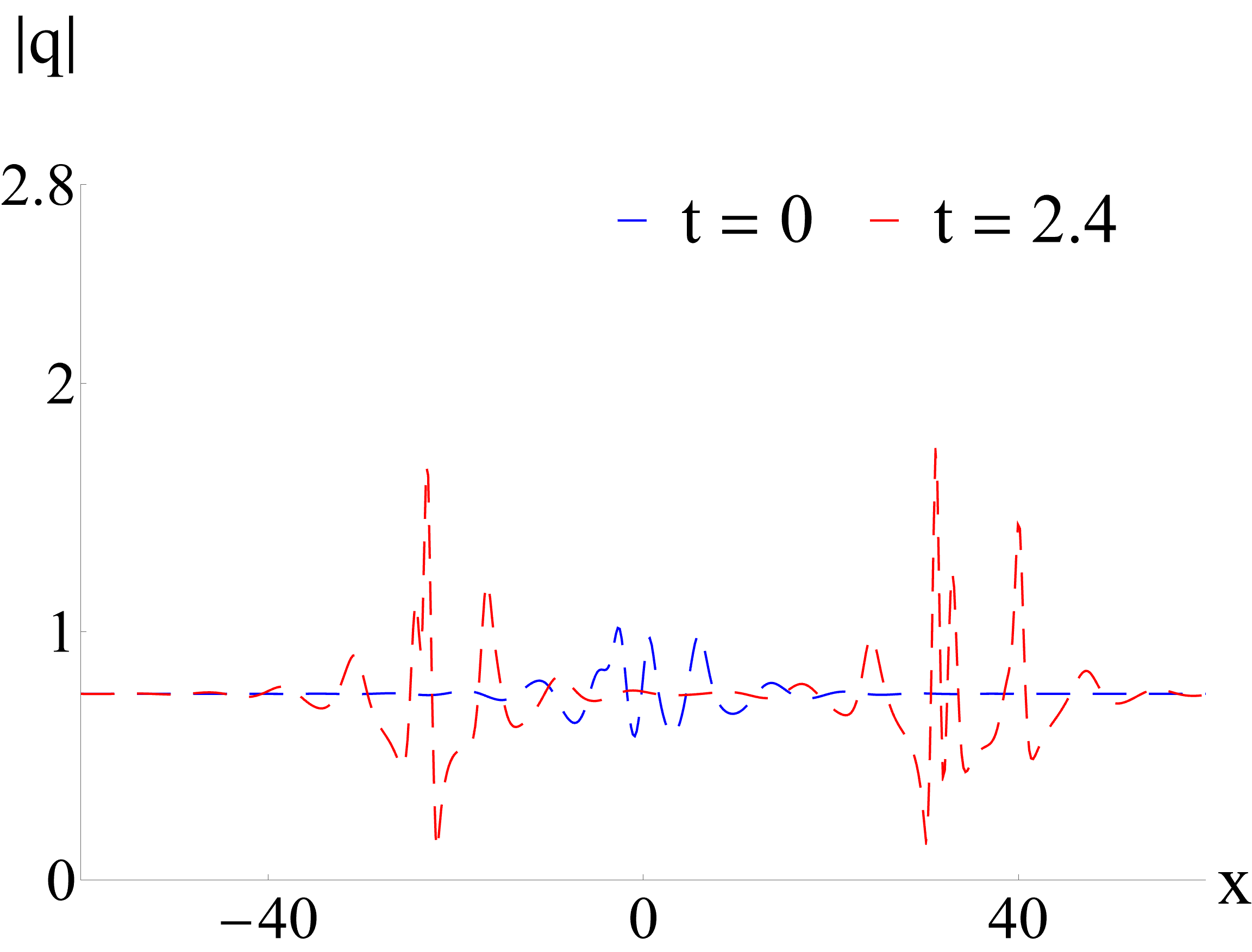}
		\end{minipage}
	\end{minipage}
	\caption{Exact twist solution density and amplitude maps. Panels (a) and (c) correspond to $|q(x, t; \chi_1,\chi_3)|$, while panels (b) and (d) correspond to $|q(x, t; \chi_2,\chi_4)|$. The parameters are $a=0.75$, $b=0.7$, $\sigma=1$, $\gamma=0.1$, $\omega=0.8$ and $x_{01}=x_{02}=t_{01}=t_{02}=0$.}
	\label{fig:niujie}
\end{figure}
Figure \ref{fig:niujie} shows twisted AB solutions. As can be seen in panels (c) and (d), these AB solutions exhibit a shape with two peaks and two troughs. As observed in the left panels, each quasi-AB solution propagates at its own group velocity. Since these velocities are determined by the corresponding eigenvalues, their variation introduces an asymmetric density profile in the SRB solution patterns. This confirms that the eigenvalue mismatch between the constituent quasi-AB solution is directly imprinted in the spatial structure of the superposition.

The above analysis can be extended to construct more complex SRB solutions involving four eigenvalues ($\chi_1$, $\chi_2$, $\chi_3$ and $\chi_4$) in the focusing regime. Such solutions can be viewed as a superposition of the structures shown in Figs. \ref{fig:jingquejie} and \ref{fig:niujie}.
\begin{figure}[htbp]
	\centering
	\noindent\begin{minipage}{\linewidth}
		\begin{minipage}[c]{0.01\linewidth} 
			\centering
			\footnotesize (a)
		\end{minipage}
		\hfill
		\begin{minipage}[c]{0.43\linewidth} 
			\centering
			\includegraphics[width=\linewidth]{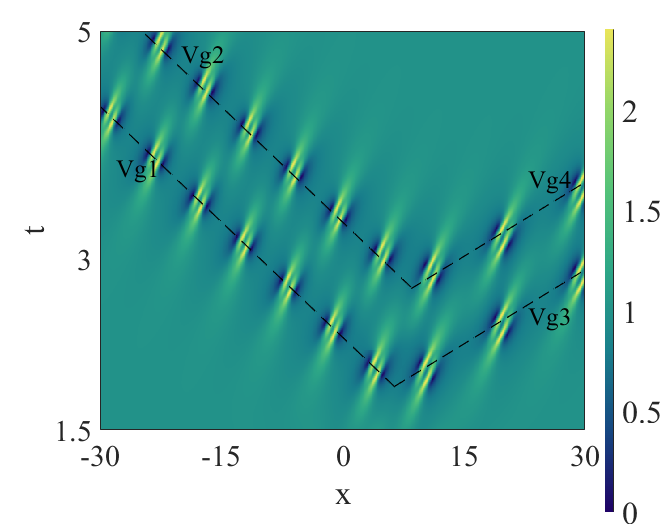}
		\end{minipage}
		\hfill
		\begin{minipage}[c]{0.01\linewidth} 
			\centering
			\footnotesize (c)
		\end{minipage}
		\hfill
		\begin{minipage}[c]{0.43\linewidth} 
			\centering
			\includegraphics[width=\linewidth]{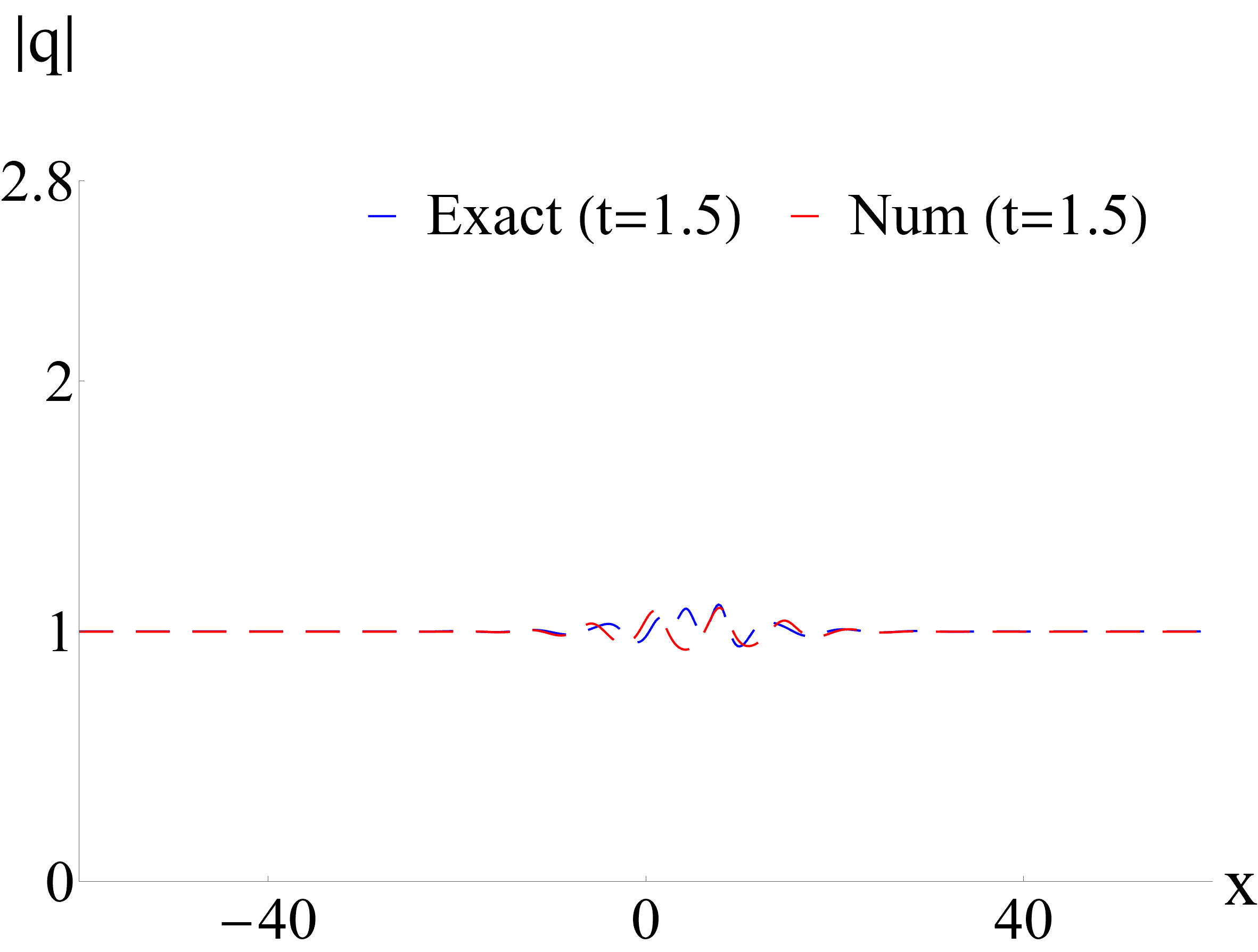}
		\end{minipage}
	\end{minipage}
	
	\vspace{0.8\baselineskip}
	
	\noindent\begin{minipage}{\linewidth}
		\begin{minipage}[c]{0.01\linewidth}  
			\centering
			\footnotesize (b)
		\end{minipage}
		\hfill
		\begin{minipage}[c]{0.43\linewidth}  
			\centering
			\includegraphics[width=\linewidth]{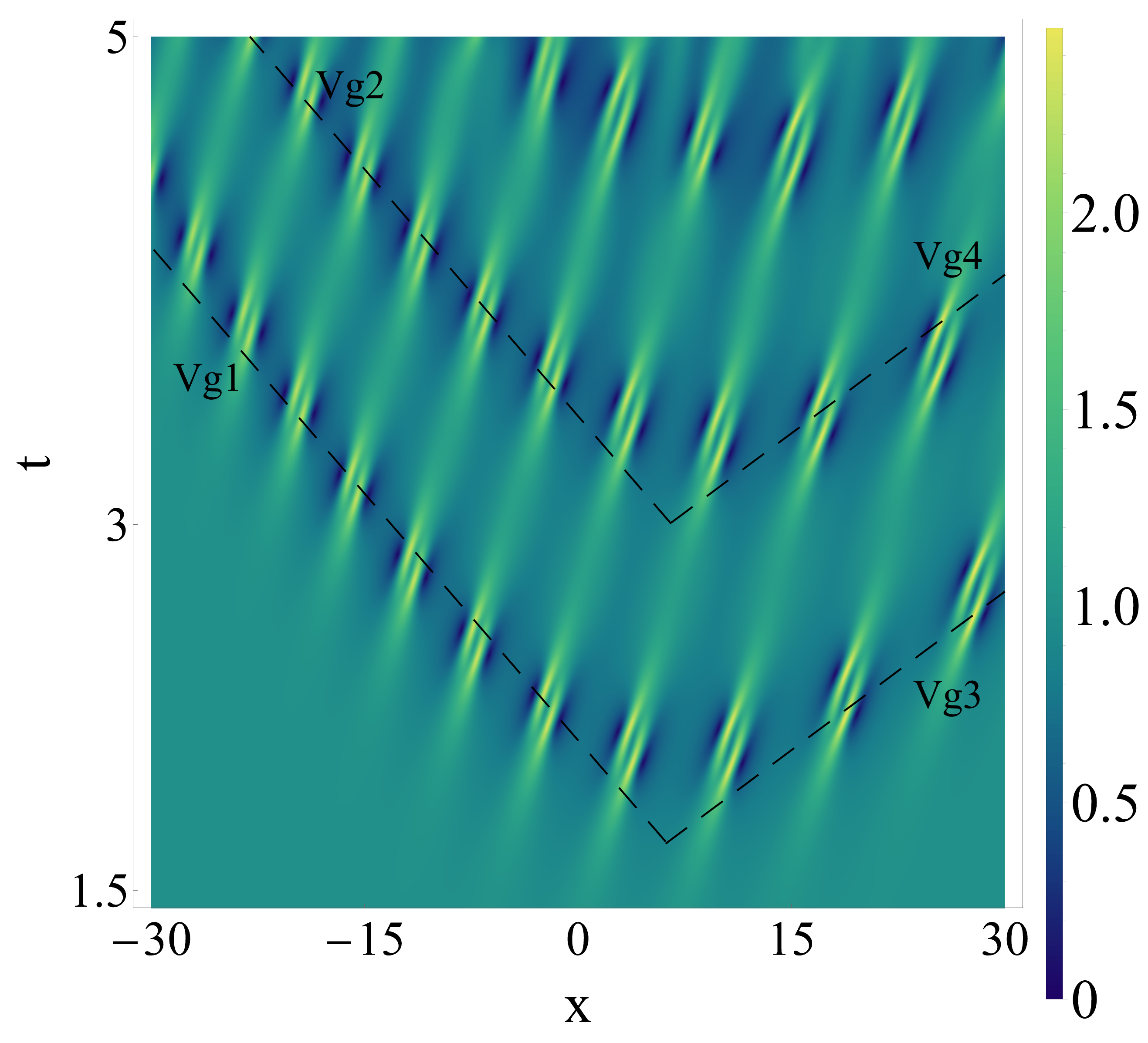}
		\end{minipage}
		\hfill
		\begin{minipage}[c]{0.01\linewidth} 
			\centering
			\footnotesize (d)
		\end{minipage}
		\hfill
		\begin{minipage}[c]{0.43\linewidth} 
			\centering
			\includegraphics[width=\linewidth]{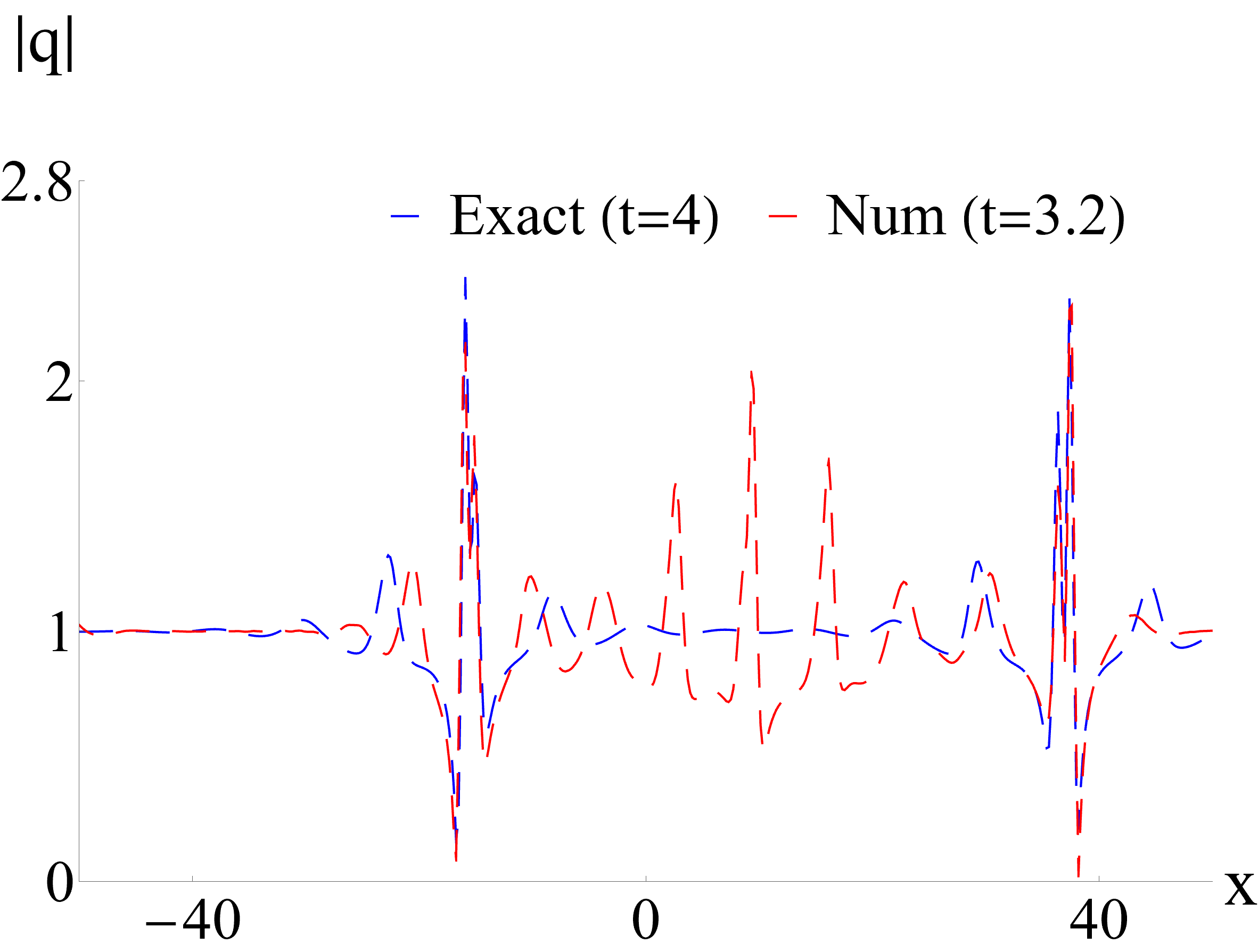}
		\end{minipage}
	\end{minipage}
	
\caption{Exact and numerical density and amplitude maps of the four-petal higher-order SRB solution. Panels (a) and (b) correspond to the exact and numerical solution $|q(x, t; \chi_1,\chi_2,\chi_3,\chi_4)|$, while panels (c) and (d) correspond to the amplitude of the solutions. The parameters are $a=1$, $b=1$, $\sigma=1$, $\gamma=0.1$, $\omega=0.8$, $\phi=3\pi/2$, $\epsilon=0.001$, $x_w=10$, $\delta x_1=\{7.238, 10.322\}$, $\delta t_1=\{1.449, 1.788\}$, $\delta x_2=\{2.450, -0.704\}$ and $\delta t_2=\{2.245, 2.083\}$.}
	\label{fig:shuzhimoni}
\end{figure}
\\

The solutions in Fig. \ref{fig:shuzhimoni} can be generated by the {localized initial condition Eq.~\eqref{eq:initial}} and the corresponding results are presented in Fig. \ref{fig:shuzhimoni}(b). Figure \ref{fig:shuzhimoni}(a) displays the second-order SRB solution of the exact solution, while Figs. \ref{fig:shuzhimoni}(c) and \ref{fig:shuzhimoni}(d) show the amplitude profiles of the two solutions at different times. 
The spatial and temporal offsets are introduced to account for the translational invariance of the SS equation.
Owing to the ongoing excitation induced by the initial condition, additional transient structures emerge during the nonlinear evolution, such as the extra structure at the top of Fig. \ref{fig:shuzhimoni}(b) and the discrepancy between the red dashed curve and the blue dashed curve in the middle portion of Fig. \ref{fig:shuzhimoni}(d).

\begin{figure}[htbp]
	\centering
	\noindent\begin{minipage}{\linewidth}
		\begin{minipage}[c]{0.01\linewidth} 
			\centering
			\footnotesize (a)
		\end{minipage}
		\hfill
		\begin{minipage}[c]{0.43\linewidth} 
			\centering
			\includegraphics[width=\linewidth]{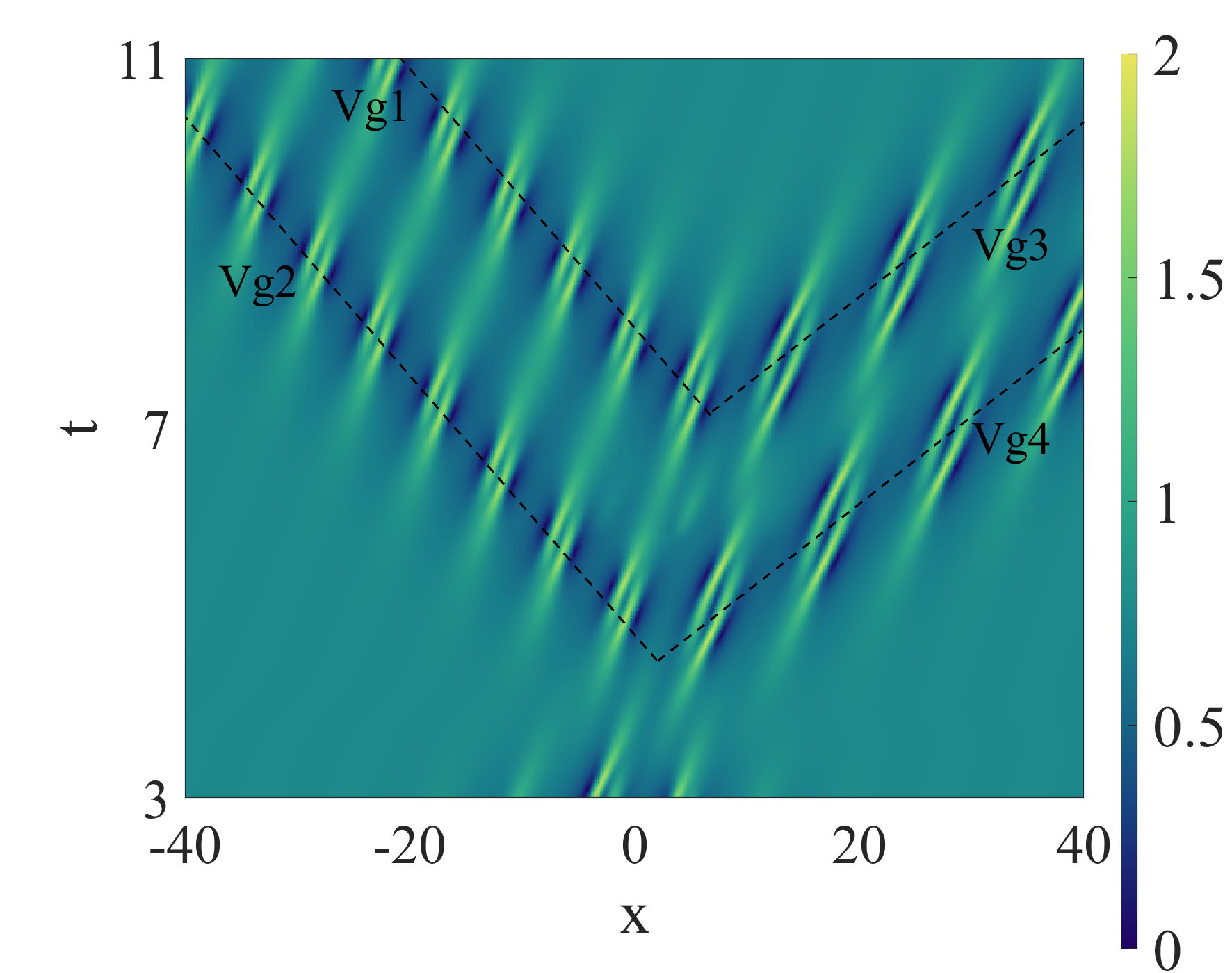}
		\end{minipage}
		\hfill
		\begin{minipage}[c]{0.01\linewidth} 
			\centering
			\footnotesize (c)
		\end{minipage}
		\hfill
		\begin{minipage}[c]{0.43\linewidth} 
			\centering
			\includegraphics[width=\linewidth]{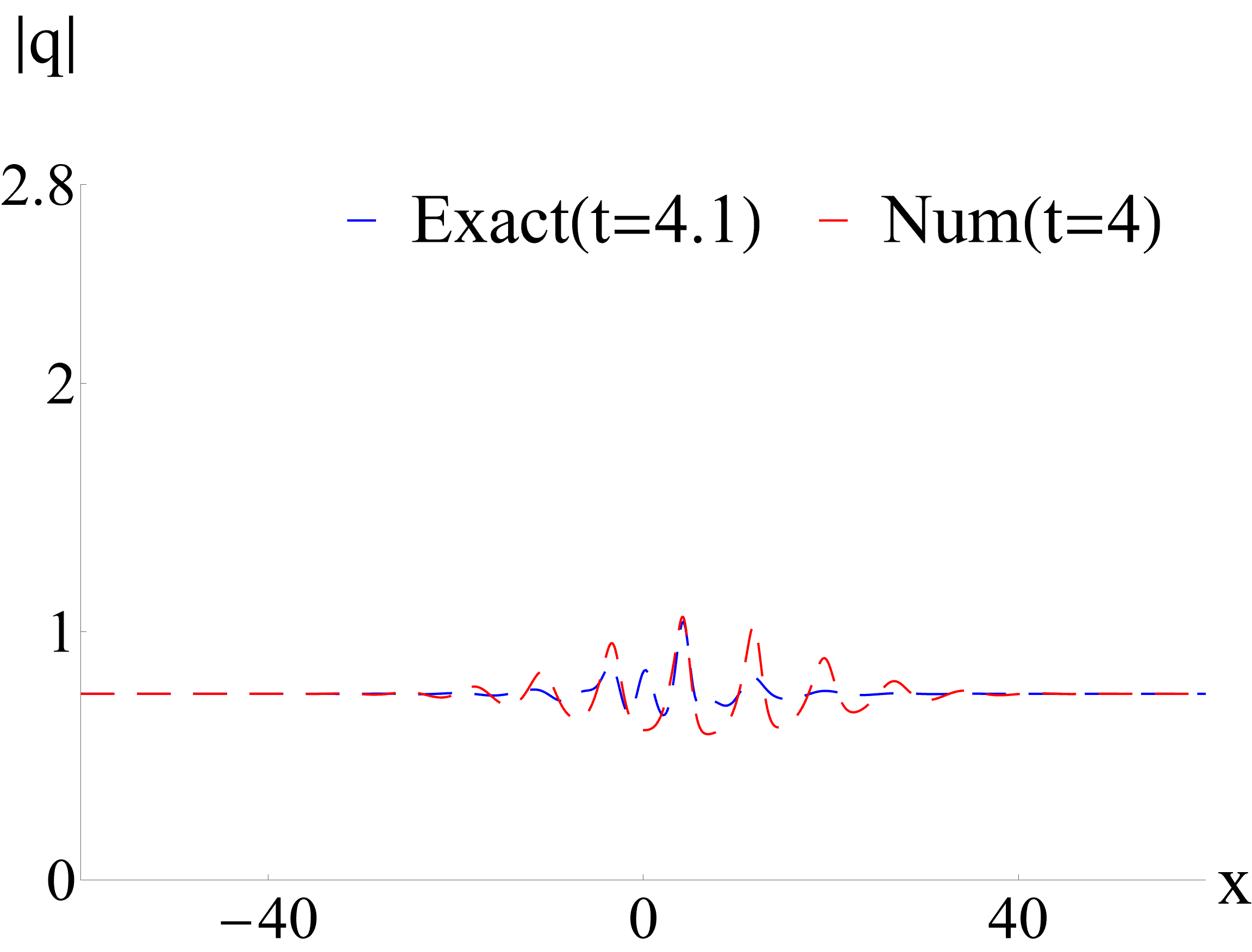}
		\end{minipage}
	\end{minipage}
	
	\vspace{0.8\baselineskip}
	
	\noindent\begin{minipage}{\linewidth}
		\begin{minipage}[c]{0.01\linewidth}  
			\centering
			\footnotesize (b)
		\end{minipage}
		\hfill
		\begin{minipage}[c]{0.43\linewidth}  
			\centering
			\includegraphics[width=\linewidth]{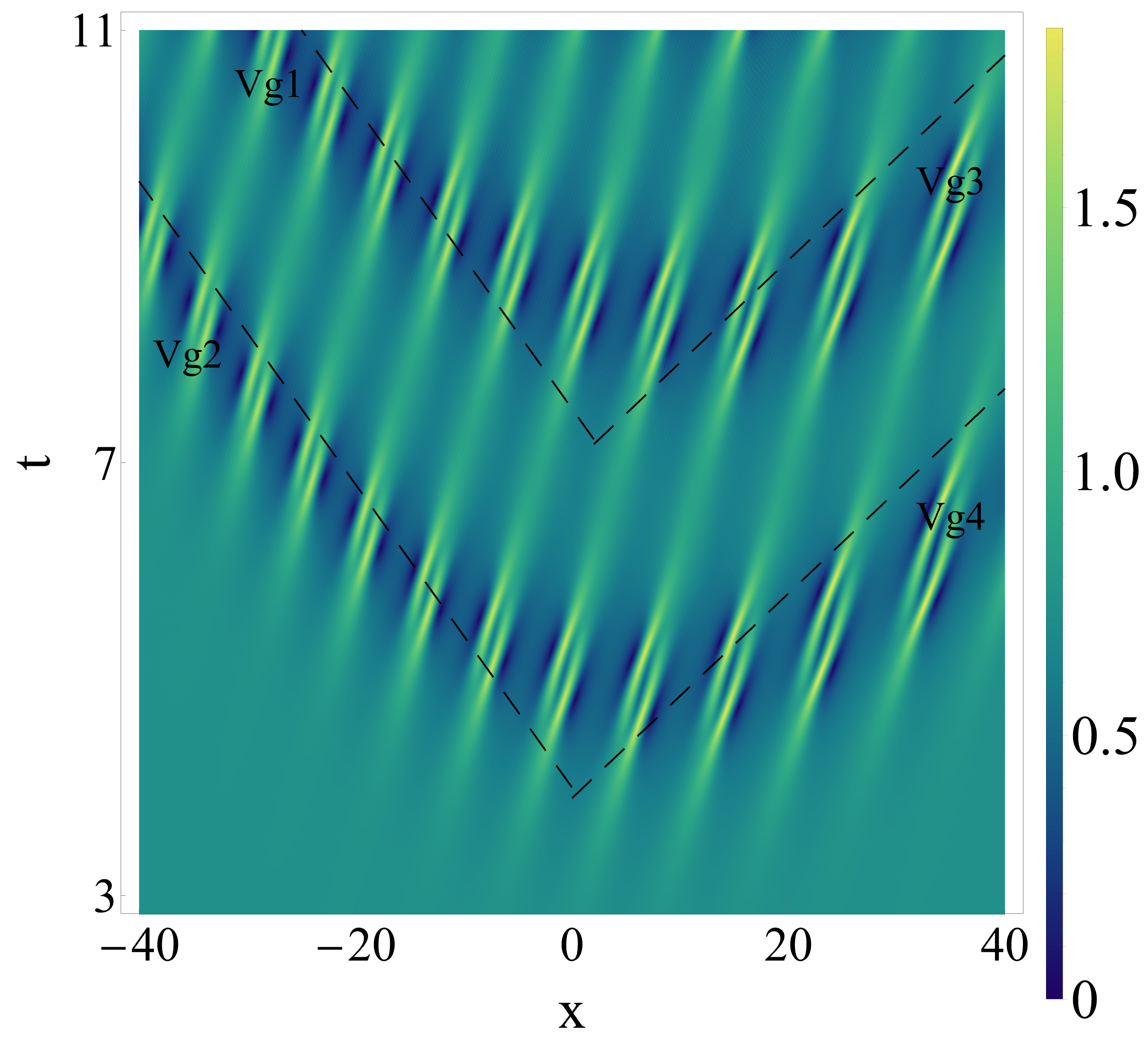}
		\end{minipage}
		\hfill
		\begin{minipage}[c]{0.01\linewidth} 
			\centering
			\footnotesize (d)
		\end{minipage}
		\hfill
		\begin{minipage}[c]{0.43\linewidth} 
			\centering
			\includegraphics[width=\linewidth]{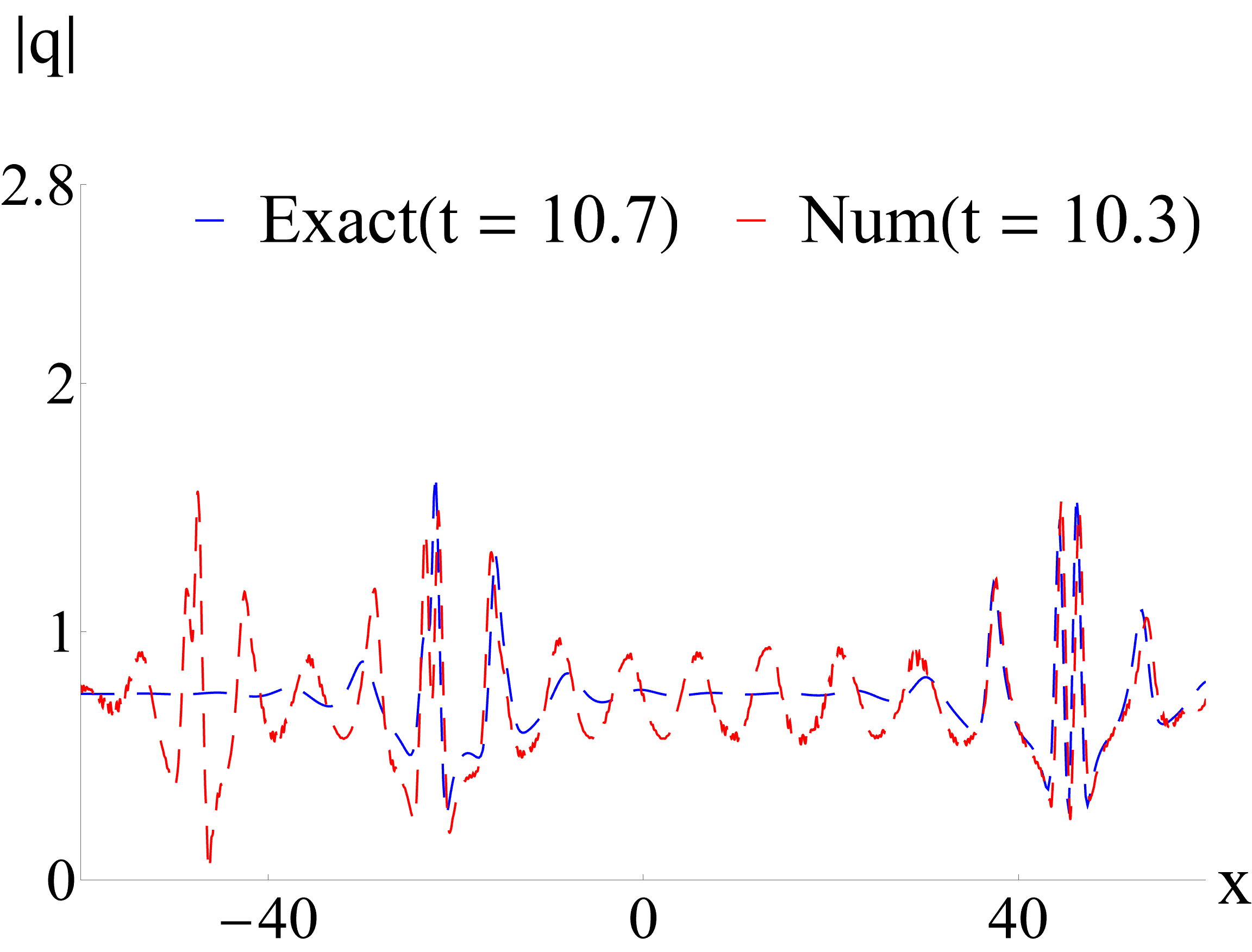}
		\end{minipage}
	\end{minipage}
	
	\caption{Exact and numerical density and amplitude maps of the twist higher-order SRB solution. Panels (a) and (b) show the density of the exact and numerical solutions $|q(x, t; \chi_1,\chi_2,\chi_3,\chi_4)|$, respectively, while panels (c) and (d) show the corresponding amplitude. The parameters are $a=0.75$, $b=0.7$, $\sigma=1$, $\gamma=0.1$, $\omega=0.8$, $\phi=3\pi/2$, $\epsilon=0.001$, $x_w=10$, $\delta x_1=\{7.238, 10.322\}$, $\delta t_1=\{1.449, 1.789\}$, $\delta x_2=\{0.328, -3.050\}$ and $\delta t_2=\{3.851, 3.223\}$.}
	\label{fig:shuzhimoni2}
\end{figure}

Figures \ref{fig:shuzhimoni2} (a) and (b) display the density distributions of the exact analytical solution and the numerically simulated solution, respectively. Panels (c) and (d) compare the corresponding amplitude profiles at different time instants, with the blue and red dashed curves representing the exact and numerical results. {At long propagation distances, successive structures partially overlap and produce the visible discrepancies. For the same $a$, $b$, and $\omega$, with $x_w\sim1/(2\gamma)$, the numerical and exact profiles differ primarily by the reported spatial and temporal shifts. Thus a finite-width perturbation reproduces the principal second-order-SRB profile without imposing a spatially periodic modulation.}

\section{Conclusion}

{For the scalar SS equation, the reduction symmetry makes the fundamental SRB construction degenerate. The paired quasi-AB branches instead produce second-order SRBs precisely in the focusing sector $n_0<0$. In the defocusing sector and in the focusing sector $n_0\geq0$, the normal-mode growth rate vanishes. The carrier wavenumber $b$ is a parameter of the selected background, rather than a coefficient of the equation, but it enters both the spectral roots and the instability interval.}

{The direct linearization gives $\mathcal{G}_{\rm lin}=3|\omega|\sqrt{-n_0}$ in the MI-active sector and hence $\mathcal{G}_{\rm lin}=|\omega\chi'|=\gamma\Delta V_g$ on the admissible branches. The formal NLS limit recovers the standard NLS gain, while the SS criterion keeps its explicit dependence on $b$. The numerical calculations show that the second-order profile is excited by localized periodic data with the predicted envelope scale. A quantitative comparison with vector Manakov dynamics requires component coupling, which is absent from the scalar model.}

\appendix

\section{Appendixes}
\subsection{Fundamental Darboux Transformation}
For the case $q=q_0$ and $\lambda=\lambda_1$, the fundamental solution to the Lax pair equations is derived by solving the linear differential equation \cite{31Guo} through the standard method
\begin{equation}
	\Phi_1=\mathbf{D}_1\mathbf{L}_1\mathbf{G_1},
\end{equation}
where
\begin{align}
\mathbf{D}_1&=\begin{bmatrix}
	1 & 0 & 0 \\
	0 & a\text{exp}(\mathrm{i}\theta_1) & 0 \\
	0 & 0 & a\text{exp}(\mathrm{i}\theta_1)
\end{bmatrix},\\\nonumber 
\theta_1 &= b \left(x + (b^2 - 6\sigma a^2) t \right),\\\nonumber
\mathbf{L}_1&=\begin{bmatrix}
	1 & 1 & 1 \\
	\frac{1}{\chi_1+b} & \frac{1}{\chi_2+b} & \frac{1}{\chi_3+b} \\
	\frac{1}{\chi_1-b} & \frac{1}{\chi_2-b} & \frac{1}{\chi_3-b}
\end{bmatrix},\\\nonumber
\mathbf{G}_1&=\begin{bmatrix}
\exp(g_1) & 0 & 0 \\
	0 & \exp(g_2) & 0 \\
	0 & 0 & \exp(g_3)
\end{bmatrix},\\\nonumber
g_i &= \mathrm{i}\chi_i\left(x + (2\lambda\chi_i + b^2 - 4\sigma a^2)t\right),
\end{align}
and $\chi_i (i = 1, 2, 3)$ are three different roots for the following cubic equation
\begin{equation}
	\label{eq:charactereq}
	\chi^3-2\lambda\chi^2-(b^2+2\sigma a^2)\chi+2 b^2\lambda=0.
\end{equation}
The fundamental DT for SS is
\begin{equation}
	\mathbf{T}=\mathbf{I}-\frac{\lambda_1-\lambda_1^*}{\lambda-\lambda_1^*}\mathbf{P}_1,\quad \mathbf{P}_1=\frac{\mathbf{\Phi}_1\mathbf{\Phi}^{\dagger}_1\mathbf{J}}{\mathbf{\Phi}^{\dagger}_1\mathbf{J}\mathbf{\Phi}_1},
\end{equation}
where $\mathbf{J}={\mathrm{diag}(1,\sigma,\sigma)}$. If we denote the eigenvector $\mathbf{\phi}_1$ as $(\phi_1,\phi_2,\phi_3)^{\mathrm{T}}$, the {B\"acklund} transformation between old potential function and new one is
\begin{equation}
q[1]=q-\frac{2\left(\lambda_1-\lambda_1^*\right)\phi_2\phi_1^*}{\left\vert \phi_1 \right\vert^2+\left\vert \phi_2 \right\vert^2+\left\vert \phi_3 \right\vert^2}.
\end{equation}
For the detailed proof, please refer to the reference \cite{32Liming}. 
\subsection{High-order Darboux Transformation}
When the spectral parameter $\lambda$ of the breather solution takes complex values, the symmetry relation is no longer satisfied. Therefore, the DT needs to be iterated and the fundamental breather solution can be obtained after twice iterated DT
\begin{align}
	\mathbf{T}_2 = \mathbf{I} - \left[ \mathbf{\Phi}_1, \mathbf{K} \mathbf{\Phi}^*_1 \right] \mathbf{G}^{-1} \mathbf{D} \begin{bmatrix} \mathbf{\Phi}^{\dagger}_1
		 \\ \mathbf{\Phi}^{\mathrm{T}}_1\mathbf{K} \end{bmatrix} \mathbf{J},
\end{align}
where 
\begin{align}
\mathbf{K}&=	\begin{bmatrix}
		-1 & 0 & 0\\0 & 0 & 1\\0 & 1 & 0
	\end{bmatrix},\quad
	\mathbf{G}=\begin{bmatrix}
		\frac{\mathbf{\Phi}^{\dagger}_1\mathbf{J}\mathbf{\Phi}_1}{\lambda_1 - \lambda_1^*} & \frac{\mathbf{\Phi}^{\dagger}_1\mathbf{J}\mathbf{K}\mathbf{\Phi}_1^*}{-2\lambda_1^*} \\[6pt]
		\frac{\mathbf{\Phi}^{\mathrm{T}}_1\mathbf{J}\mathbf{K}\mathbf{\Phi}_1}{2 \lambda_1} & \frac{\mathbf{\Phi}^{\dagger}_1\mathbf{J}\mathbf{\Phi}_1}{\lambda_1 - \lambda_1^*}
	\end{bmatrix}, \\\nonumber
	\mathbf{D}&= \text{diag} \left( \frac{1}{\lambda - \lambda_1^*}, \frac{1}{\lambda + \lambda_1} \right).
\end{align}
Based on this, the N-fold Darboux transformation can be obtained. For details, see \cite{32Liming}.

Suppose we have $n$ different solutions $ \mathbf{\Phi}_k = (\phi_{1,k}, \phi_{2,k}, \phi_{3,k})^{\mathrm{T}} $ for system \eqref{eq:lax} at $ \lambda = \lambda_k \in \mathrm{i}\mathbb{R} $ such that $ \phi_{2,k} \phi_{1,k}^* + \phi_{3,k}^* \phi_{1,k} = 0 $ ($ k = 1, 2, \dots, n $) and $ m \equiv N - n $ different solutions $ \mathbf{\Psi_l} $ for system \eqref{eq:lax} at $ \lambda = \mu_l \notin \mathrm{i}\mathbb{R} $ ($ l = 1, 2, \dots, m $), then we have the following  N-fold  DT 
\begin{equation}
    \mathbf{T_N} = \mathbf{I} - \mathbf{Y} \mathbf{G}^{-1} \mathbf{D} \mathbf{Y}^\dagger \mathbf{J},
\end{equation}
where
\begin{align*}
	\mathbf{Y} &= \begin{bmatrix} 
		\mathbf{\Phi}_1, & \mathbf{\Phi}_2, & \dots, & \mathbf{\Phi}_n, & \mathbf{\Psi}_1, \mathbf{K}\mathbf{\Psi}_1^*, & \dots, & \mathbf{\Psi}_m, & \mathbf{K}\mathbf{\Psi}_m^*
	\end{bmatrix},\\[1.2ex]
	\mathbf{D} &= \operatorname{diag} \left(
	\frac{1}{\lambda - \lambda^*_1}, \frac{1}{\lambda - \lambda^*_2}, \dots, \frac{1}{\lambda - \lambda^*_n},
	\frac{1}{\lambda - \mu^*_1}, \frac{1}{\lambda + \mu_1}, \right. \nonumber \\[0.6ex] &\qquad \left.
	\,\,\,\,\,\,\,\,\,\,\,\,\,\,\frac{1}{\lambda - \mu^*_2}, \frac{1}{\lambda + \mu_2}, \dots,
	\frac{1}{\lambda - \mu^*_m}, \frac{1}{\lambda + \mu_m} 
	\right),\\[1.5ex]
	\mathbf{G} &= \begin{bmatrix} 
		\mathbf{G}_{11} & \mathbf{G}_{12} \\ 
		-\mathbf{G}_{12}^\dagger & \mathbf{G}_{22} 
	\end{bmatrix},\\[1.2ex]
	\mathbf{G}_{11} &= \begin{bmatrix} 
		\dfrac{\mathbf{\Phi}^{\dagger}_1 \mathbf{J} \mathbf{\Phi_1}}{\lambda_1 - \lambda^*_1} & \cdots & \dfrac{\mathbf{\Phi}^{\dagger}_1 \mathbf{J} \mathbf{\Phi_n}}{\lambda_n - \lambda^*_1} \\[1.2ex]
		\vdots & \ddots & \vdots \\[1.2ex]
		\dfrac{\mathbf{\Phi}^{\dagger}_n \mathbf{J} \mathbf{\Phi_1}}{\lambda_1 - \lambda^*_n} & \cdots & \dfrac{\mathbf{\Phi}^{\dagger}_n \mathbf{J} \mathbf{\Phi_n}}{\lambda_n - \lambda^*_n} 
	\end{bmatrix},\\\\
	\mathbf{G}_{12} &= \begin{bmatrix} 
		\dfrac{\mathbf{\Phi}^{\dagger}_1 \mathbf{J} \mathbf{\Psi}_1}{\mu_1 - \lambda^*_1} & 
		\dfrac{\mathbf{\Phi}^{\dagger}_1 \mathbf{J}\mathbf{K} \mathbf{\Psi}_1^*}{-\mu^*_1 - \lambda^*_1} & \cdots & 
		\dfrac{\mathbf{\Phi}^{\dagger}_1 \mathbf{J} \mathbf{\Psi}_m}{\mu_m - \lambda^*_1} & 
		\dfrac{\mathbf{\Phi}^{\dagger}_1 \mathbf{J}\mathbf{K} \mathbf{\Psi}_m^*}{-\mu^*_m - \lambda^*_1} \\[1.2ex]
		\vdots & \vdots & \ddots & \vdots & \vdots \\[1.2ex]
		\dfrac{\mathbf{\Phi}^{\dagger}_n \mathbf{J} \mathbf{\Psi}_1}{\mu_1 - \lambda^*_n} & 
		\dfrac{\mathbf{\Phi}^{\dagger}_n \mathbf{J}\mathbf{K} \mathbf{\Psi}_1^*}{-\mu^*_1 - \lambda^*_n} & \cdots & 
		\dfrac{\mathbf{\Phi}^{\dagger}_n \mathbf{J} \mathbf{\Psi}_m}{\mu_m - \lambda^*_n} & 
		\dfrac{\mathbf{\Phi}^{\dagger}_n \mathbf{J}\mathbf{K} \mathbf{\Psi}_m^*}{-\mu^*_m - \lambda^*_n} 
	\end{bmatrix},\\[1.5ex]
	\mathbf{G}_{22} &= \begin{bmatrix} 
		\dfrac{\mathbf{\Psi}^{\dagger}_1 \mathbf{J} \mathbf{\Psi}_1}{\mu_1 - \mu^*_1} & 
		\dfrac{\mathbf{\Psi}^{\dagger}_1 \mathbf{J}\mathbf{K} \mathbf{\Psi}_1^*}{-\mu^*_1 - \mu^*_1} & \cdots & 
		\dfrac{\mathbf{\Psi}^{\dagger}_1 \mathbf{J} \mathbf{\Psi}_m}{\mu_m - \mu^*_1} & 
		\dfrac{\mathbf{\Psi}^{\dagger}_1 \mathbf{J}\mathbf{K} \mathbf{\Psi}_m^*}{-\mu^*_m - \mu^*_1} \\[2ex]
		\dfrac{\mathbf{\Psi}^{\mathrm{T}}_1 \mathbf{J}\mathbf{K} \mathbf{\Psi}_1}{\mu_1 + \mu_1} & 
		\dfrac{\mathbf{\Psi}^{\mathrm{T}}_1 \mathbf{J} \mathbf{\Psi}_1^*}{-\mu^*_1 + \mu_1} & \cdots & 
		\dfrac{\mathbf{\Psi}^{\mathrm{T}}_1 \mathbf{J}\mathbf{K} \mathbf{\Psi}_m}{\mu_m + \mu_1} & 
		\dfrac{\mathbf{\Psi}^{\mathrm{T}}_1 \mathbf{J} \mathbf{\Psi}_m^*}{-\mu^*_m + \mu_1}  
		\\[1.2ex]
		\vdots & \vdots & \ddots & \vdots & \vdots \\[1.2ex]
		\dfrac{\mathbf{\Psi}^{\dagger}_m \mathbf{J} \mathbf{\Psi}_1}{\mu_1 - \mu^*_m} & 
		\dfrac{\mathbf{\Psi}^{\dagger}_m \mathbf{J}\mathbf{K} \mathbf{\Psi}_1^*}{-\mu^*_1 - \mu^*_m} & \cdots & 
		\dfrac{\mathbf{\Psi}^{\dagger}_m \mathbf{J} \mathbf{\Psi}_m}{\mu_m - \mu^*_m} & 
		\dfrac{\mathbf{\Psi}^{\dagger}_m \mathbf{J}\mathbf{K} \mathbf{\Psi}_m^*}{-\mu^*_m - \mu^*_m} \\[2ex]
		\dfrac{\mathbf{\Psi}^{\mathrm{T}}_m \mathbf{J}\mathbf{K} \mathbf{\Psi}_1}{\mu_1 + \mu_m} & 
		\dfrac{\mathbf{\Psi}^{\mathrm{T}}_m \mathbf{J}\mathbf{\Psi}_1^*}{-\mu^*_1 + \mu_m} & \cdots & 
		\dfrac{\mathbf{\Psi}^{\mathrm{T}}_m \mathbf{J}\mathbf{K} \mathbf{\Psi}_m}{\mu_m + \mu_m} & 
		\dfrac{\mathbf{\Psi}^{\mathrm{T}}_m \mathbf{J}\mathbf{\Psi}_m^*}{-\mu^*_m + \mu_m}  
	\end{bmatrix}.
\end{align*}
And the B\"acklund transformation for SS\eqref{eq:eq} equation can be represented as
\begin{equation}
	q[N] = q - 1 + \frac{\det(\mathbf{G} - 2\mathbf{Y}
		_1^\dagger \mathbf{Y}_2)}{\det(\mathbf{G})},
\end{equation}
where $ \mathbf{Y}_i $ represents the $ i $-th row of matrix $ \mathbf{Y} $.

When $N=2$, we can obtain the fundamental solution of Eq. \eqref{eq:eq}. Therefore, we have 
\begin{equation}
	\label{eq:qb}
	q_b=\frac{\delta_1 \delta_2}{\delta_1\delta_1^* + a^2 \delta_2\delta_2^* +a^2 \delta_3\delta_3^*},
\end{equation}
where
\begin{align*}
	\delta_1 =\lambda_r&  \Big( (1+\mu(\chi)) (e^{\tau} + e^{\mathrm{i}\Lambda}) \\\nonumber
	&+(1+\nu(\chi)) (e^{-\mathrm{i}\Lambda} + e^{-\tau}) \\\nonumber
	&+ (1+\nu(\chi^*)) (e^{\mathrm{i}\Lambda} + e^{-\tau} e^{2\mathrm{i}\Lambda})\\\nonumber 
	&+ (1+\mu(\hat{\chi})) (e^{-\tau} + e^{-2\tau} e^{\mathrm{i}\Lambda}) \Big) \\\nonumber
	- \mathrm{i}\lambda_i& \Big( (H(\chi) -\mu(\chi)) e^{\tau} 
	+ (H(\chi) -\nu(\chi)) e^{-\mathrm{i}\Lambda} \\\nonumber
	&+ (G(\hat{\chi})+ G(\chi) -\mu(\chi)-\nu(\chi^*) ) e^{\mathrm{i}\Lambda}\\\nonumber
	&+ (G(\hat{\chi})+ G(\chi)-\nu(\chi)-\mu(\hat{\chi})) e^{-\tau} \\\nonumber
	&+ (H(\hat{\chi})-\nu(\chi^*)) e^{-\tau} e^{2\mathrm{i}\Lambda} 
	+ (H(\hat{\chi})-\mu(\hat{\chi})) e^{-2\tau} e^{\mathrm{i}\Lambda} \Big),\\\nonumber
	\delta_2=  \lambda_r& \Big( (1+\mu(\chi)) \left(F(-\chi^*)e^{\tau}
	+ F(-\hat{\chi}^*)e^{-\mathrm{i}\Lambda} \right)\\\nonumber 
	&+(1+\nu(\chi)) \left( F(-\chi^*) e^{-\mathrm{i}\Lambda}
	+ F(-\hat{\chi}^*)e^{-\tau} e^{-2\mathrm{i}\Lambda} \right) \\ \nonumber
	& + (1+\nu(\chi^*)) \left(F(-\chi^*) e^{\mathrm{i}\Lambda}
	+ F(-\hat{\chi}^*)e^{-\tau} \right)\\\nonumber
	&+ (1+\mu(\hat{\chi})) \left(F(-\chi^*) e^{-\tau}
	+ F(-\hat{\chi}^*)e^{-2\tau} e^{-\mathrm{i}\Lambda} \right) \Big)\\ \nonumber
	+ \mathrm{i}\lambda_i& \Big( \left(  F(\chi) + (1+M(\chi))\tilde{F}(\chi^*) \right) e^{\tau} \\\nonumber
	&+ \left( F(\chi) + (1+M(\chi))\tilde{F}(\hat{\chi^*}) \right) e^{-\mathrm{i}\Lambda}\\\nonumber
	&+ \left(  F(\chi) + (1+N(\chi))\tilde{F}(\chi^*) \right) e^{-\mathrm{i}\Lambda}\\\nonumber 
	&+ \left(  F(\chi) + (1+N(\chi))\tilde{F}(\hat{\chi^*}) \right) e^{-\tau} e^{-2\mathrm{i}\Lambda} \\\nonumber
	&	+ \left( F(\hat{\chi}) + (1+N(\hat{\chi}))\tilde{F}(\chi^*) \right) e^{\mathrm{i}\Lambda}\\\nonumber 
	&+ \left( F(\hat{\chi}) + (1+N(\hat{\chi}))\tilde{F}(\hat{\chi^*}) \right) e^{-\tau}\\\nonumber
	&+ \left( F(\hat{\chi}) + (1+M(\hat{\chi}))\tilde{F}(\chi^*) \right) e^{-\tau} \\\nonumber
	&+ \left( F(\hat{\chi}) + (1+M(\hat{\chi}))\tilde{F}(\hat{\chi^*}) \right) e^{-2\tau} e^{-\mathrm{i}\Lambda} \Big),\\\nonumber
	\delta_3= \lambda_r& \Big( (1+\mu(\chi))\left(\tilde{F}(-\chi^*)e^{\tau}  +  \tilde{F}(-\hat{\chi^*})e^{-\mathrm{i}\Lambda}\right)\\\nonumber&+(1+\nu(\chi^*))\left( \tilde{F}(-\chi^*)e^{\mathrm{i}\Lambda} + \tilde{F}(-\hat{\chi^*})e^{-\tau}) \right)\\\nonumber
	& +(1+\nu(\chi))\left( \tilde{F}(-\chi^*)e^{-\mathrm{i}\Lambda} + \tilde{F}(-\hat{\chi^*}) e^{-\tau}e^{-2\mathrm{i}\Lambda}\right)\\\nonumber
	&+(1+\mu(\hat{\chi}))\left(\tilde{F}(-\chi^*)e^{-\tau}  + \tilde{F}(-\hat{\chi^*})) e^{-2\tau}e^{-\mathrm{i}\Lambda}\right)\Big)\\\nonumber
	+ \mathrm{i}\lambda_i& \Big( \left(\tilde{F}(\chi) + (1-M(\chi))F(\chi^*) \right) e^{\tau}\\\nonumber 
	&+ \left( \tilde{F}(\chi) + (1-M(\chi))F(\hat{\chi^*}) \right) e^{-\mathrm{i}\Lambda}\\ \nonumber
	&	+ \left( \tilde{F}(\chi) +(1-N(\chi))F(\chi^*) \right) e^{-\mathrm{i}\Lambda}\\\nonumber& 
	+ \left( \tilde{F}(\chi) +(1-N(\chi))F(\hat{\chi^*}) \right) e^{-\tau} e^{-2\mathrm{i}\Lambda} \\\nonumber
	&	+ \left( \tilde{F}(\hat{\chi}) + (1-N(\hat{\chi}))F(\chi^*) \right) e^{\mathrm{i}\Lambda}\\\nonumber& 
	+ \left( \tilde{F}(\hat{\chi}) + (1-N(\hat{\chi}))F(\hat{\chi^*}) \right) e^{-\tau} \\\nonumber
	& + \left( \tilde{F}(\hat{\chi}) +(1-M(\hat{\chi}))F(\chi^*) \right) e^{-\tau} \\\nonumber&
	+ \left( \tilde{F}(\hat{\chi}) +(1-M(\hat{\chi}))F(\hat{\chi^*}) \right) e^{-2\tau} e^{-\mathrm{i}\Lambda} \Big),
\end{align*}
with 
\begin{align*}
	G(\chi)&=\sum_{i=1}^{2}\frac{a^2}{(-\chi+b_i)(\hat{\chi}+b_i)},\\ H(\chi)&=\sum_{i=1}^{2}\frac{a^2}{(-\chi+b_i)(\chi+b_i)},\\
	M(\chi)&=\frac{a^2}{(-\chi^*-b)(\chi+b)}-\frac{a^2}{(-\chi^*+b)(\chi-b)},\\
	N(\chi)&=\frac{a^2}{(-\hat{\chi}^*-b)(\chi+b)}-\frac{a^2}{(-\hat{\chi}^*+b)(\chi-b)}.
\end{align*}


\bibliography{reference}

\end{document}